\documentclass[reprint, floatfix,eqsecnum,superscriptaddress,preprint,nofootinbib,onecolumn,amsmath,amssymb,aps,prb]{revtex4-2}
\pdfoutput=1

\usepackage{graphicx}
\usepackage{dcolumn}
\usepackage{bm}
\usepackage{hyperref}
\usepackage{braket}
\usepackage{shorthand}
\usepackage[caption=false]{subfig}
\usepackage{makecell}
\usepackage{xcolor}
\usepackage[normalem]{ulem}

\newcommand{\tchi}{\tilde{\chi}}

\begin{document}

\begin{titlepage}

\title{Thermalization of Randomly Coupled SYK Models}

\author{Ramanjit Sohal}
\affiliation{ Department of Physics, Princeton University, Princeton, New Jersey, 08540, USA}
\affiliation{Department of Physics and Institute for Condensed Matter Theory,
University of Illinois at Urbana-Champaign, 1110 West Green Street, Urbana, Illinois 61801, USA}
\author{Laimei Nie}
\affiliation{Department of Physics and Institute for Condensed Matter Theory,
University of Illinois at Urbana-Champaign, 1110 West Green Street, Urbana, Illinois 61801, USA}
\author{Xiao-Qi Sun}
\affiliation{Department of Physics and Institute for Condensed Matter Theory,
University of Illinois at Urbana-Champaign, 1110 West Green Street, Urbana, Illinois 61801, USA}
\author{Eduardo Fradkin}
\affiliation{Department of Physics and Institute for Condensed Matter Theory,
University of Illinois at Urbana-Champaign, 1110 West Green Street, Urbana, Illinois 61801, USA}

\date{\today}

\begin{abstract}
We investigate the thermalization of Sachdev-Ye-Kitaev (SYK) models coupled via random interactions following quenches from the perspective of entanglement.
Previous studies have shown that when a system of two SYK models coupled by random two-body terms
is quenched from the thermofield double state with sufficiently low effective temperature, the R\'enyi entropies do \emph{not} saturate to the expected thermal values in the large-$N$ limit. Using numerical large-$N$ methods, we first show that the R\'enyi entropies in a pair SYK models coupled by two-body terms can thermalize, if quenched from a state with sufficiently high effective temperature, and hence exhibit state-dependent thermalization. In contrast, SYK models coupled by single-body terms appear to always thermalize.
We provide evidence that the subthermal behavior in the former system is likely a large-$N$ artifact by repeating the quench for finite $N$ and finding that the saturation value of the R\'enyi entropy extrapolates to the expected thermal value in the $N \to \infty$ limit. Finally, as a finer grained measure of thermalization, we compute the late-time spectral form factor of the reduced density matrix after the quench. While a single SYK dot exhibits perfect agreement with random matrix theory, both the quadratically and quartically coupled SYK models exhibit slight deviations. 
\end{abstract}

\maketitle

\end{titlepage}


\section{Introduction}

The Sachdev-Ye-Kitaev (SYK) model \cite{Sachdev1993,Kitaev2015,Maldacena2016} has emerged in recent years as a 
rare example of an exactly solvable (in a large-$N$ limit) chaotic quantum many-body system.
Indeed, this toy model, comprising $N$ Majorana fermions interacting via random all-to-all interactions, exhibits spectral statistics largely consistent with those predicted by random matrix theory \cite{garciagarcia2016,You2017,Cotler2017} and out-of-time-order correlators displaying exponential decay with Lyapunov exponent saturating the chaos bound in the large-$N$ limit \cite{Kitaev2015,Maldacena2016}, thereby satisfying both standard and more modern criteria for quantum chaos. 
These remarkable features have prompted the study of a slew of generalizations of the SYK model, constructed by coupling together multiple SYK systems, which we will refer to as SYK ``dots" \cite{Chen2017,Banerjee2017,Haldar2018,milekhin2021nonlocal}. Such coupled SYK models exhibit a rich phenomenology, depending on the type of interdot coupling, ranging from the realization of phases of matter holographically dual to wormholes \cite{Maldacena2018} to interesting symmetry breaking phases \cite{Kim2019,Klebanov2020,Sahoo2020}. Moreover, the coupling of multiple SYK dots with suitably local interactions allows for the construction of SYK ``chains" and hence the study of chaos in higher dimensions \cite{Gu2017a,Gu2017c,Song2017,Jian2017,Davison2017,Jian2018,Khevschenko2018,Khveshchenko2020,Liu2020}.

A key feature of such chaotic systems is that they are expected to rapidly thermalize when taken out of equilibrium.
This is formalized by the eigenstate thermalization hypothesis (ETH) \cite{Deutsch1991,Srednicki1994,DAlessio2016}, which, loosely speaking, implies that generic quantum systems, when quenched from an initial state, should equilibrate to a ``thermal" state, in which observables assume expectation values equal to those in a thermal ensemble with temperature set by the energy density. In particular, one expects the subsystem entanglement entropy to saturate to the thermal entropy at this effective temperature. Several studies have characterized such non-equilibrium dynamics of the SYK model and its variants \cite{Eberlein2017,Gu2017b,almheiri2019universal,Chen2020,Zhang2020b,Liu2018,Zhang2019,Bhattacharya2019,Haldar2020,Jian2021,Samui2021,Larzul2021}. Of particular interest for us is the surprising result of Ref. \cite{Gu2017b}, in which the authors 
studied an SYK chain formed by coupling SYK dots with random two-body interaction in the large-$N$ limit \cite{Gu2017a}.
When quenched from thermofield double states with low effective temperature and in the limit of weak interdot coupling, the R\'enyi entropies were found to \emph{not} saturate to the expected thermal values, even in the special case of a chain with only two sites. This subthermal behavior suggests a pair of SYK dots coupled in this way does not rapidly thermalize. This is surprising, given that this SYK chain was found to have maximal Lyapunov exponent \cite{Gu2017a} and, moreover, the expectation values of local few-body operators in the SYK model and its variants, including this SYK chain, have been shown to be consistent with ETH in finite-size systems \cite{Sonner2017,HunterJones2018,Halataei2021}. It was conjectured in Ref. \cite{Gu2017b}, however, that this slow thermalization was an artifact of the large-$N$ and low effective temperature limits, which could result in certain ``heavy" modes taking an infinite time to thermalize. We note that the von Neumann entropy is expected to be insensitive to the tail effects in the entanglement spectrum induced by such modes and hence to thermalize \cite{Penington2019}. While one may conclude this coupled SYK model is thermalizing by this measure, it is still of interest to better understand the validity of this picture of non-thermalizing heavy modes and the non-equilibrium dynamics of the R\'enyi entropies, which are more sensitive to outliers in the entanglement spectrum.

This sets the context for the present work, the goal of which is to provide a more detailed characterization of thermalization, or the lack thereof, in coupled SYK models, both in the large-$N$ limit and for finite $N$. In particular, we consider both the two-site version of the SYK chain, comprising two SYK dots coupled by random interactions quartic in fermion operators, as well as a model of two SYK dots coupled by random quadratic interactions for comparison. In addition to reconsidering the late-time post-quench entanglement, we characterize chaos and thermalization in these models through complementary diagnostics including an analysis of the late-time entanglement spectrum. Before delving into our analysis, we first provide an overview of the strategies employed and our results.

\subsection{Summary of Results}
First, we study quenches from high-energy pure states -- 
focusing on the Kourkoulou-Maldacena states \cite{Kourkoulou2017} -- in the large-$N$ limit of both coupled SYK models. 
We numerically compute the R\'enyi entropy  exactly
, allowing us to consider larger interdot couplings and pure states with larger effective temperatures than those considered in Ref. \cite{Gu2017b}.
We illustrate that for sufficiently large coupling strength, the quartically coupled model in fact exhibits \emph{state-dependent} thermalization: the R\'enyi entropy saturates to the expected thermal value for states with sufficiently high effective temperature and plateaus at a subthermal value for lower effective temperatures. 
In contrast, at least for the parameter regimes considered, the quadratically coupled model always thermalizes. 
We also consider quenches from thermofield double states in Appendix \ref{app:thermofield-double}, for which we also find state-dependent thermalization.

In order to better understand the extent to which these effects are artifacts of the large-$N$ limit and provide more fine-grained measures of thermalization, we next study these coupled models for small values of $N$ using exact diagonalization (ED). Considering first their spectral properties, we find that the spectral form factors (SFFs) of both coupled models, which provide a measure of spectral rigidity, agree well with random matrix theory (RMT) predictions. Additionally, the eigenstate entanglement entropies of the two models appear consistent with ETH.
Moreover, we find that the R\'enyi and entanglement entropies saturate to the expected thermal values after quenches at finite $N$ in both coupled models. 
Indeed, when extrapolated to $N \to \infty$, the finite-$N$ numerics appear consistent with thermalization, suggesting that the subthermal behavior in the large-$N$ limit is indeed an artifact of said limit.

We further scrutinize the late-time states obtained after these quenches by computing the SFF of the late-time reduced density matrix (RDM) \cite{Chen2018,Chang2019} in both the coupled SYK models and the regular SYK model, for comparison. 
Heuristically, assuming the form of ETH proposed in Ref. \cite{Garrison2018} holds and the RDM resembles a thermal density matrix, then its singular values should exhibit a spectral rigidity similar to that of the physical energy spectrum,
providing another means of characterizing thermalization.
Indeed, we find that the RDM SFF after a quench in the standard SYK model exhibits perfect agreement with RMT predictions, while both coupled SYK models exhibit slight deviations. Our analysis yields a novel perspective on the maximally chaotic nature of the single SYK model and provides further evidence that the physical origin of the quartically coupled model's subthermal behavior does not persist down to finite-$N$. We are thus led to conclude that the quartically coupled model behaves as a generic non-integrable system at finite-$N$ and only exhibits state-dependent thermalization as a large-$N$ artifact. 

The remainder of this manuscript is structured as follows. In Section \ref{sec:models-and-quench} we introduce the models under consideration and the quenches we will perform. In Section \ref{sec:large-N}, 
we review the path integral setup for evaluating the R\'enyi entropy, derive self-consistent equations for the R\'enyi entropy for these models in the large-$N$ limit, and present our numerical results.  
We proceed to our finite $N$ analysis of the coupled SYK models, first investigating the spectral properties in Section \ref{sec:spectral-properties}, followed by results on the entanglement entropies after a quench as well as RDM SFF of each model in Section \ref{sec:finite-N-quenches}.
Finally, we discuss our results and conclude in Section \ref{sec:discussion}.

\section{Models and Quench Set-Up \label{sec:models-and-quench} }

\subsection{Models}

The SYK model \cite{Kitaev2015,Maldacena2016} is a zero-dimensional system of $N$ Majorana fermions coupled via random all-to-all interactions, as described by the Hamiltonian
\begin{align}
	H_q &= \sum_{1 \leq i_1 < \dots < i_q \leq N} i^{q/2} J_{i_1\dots i_q} \chi_{i_1} \dots \chi_{i_q}.
\end{align}
Here, the couplings are sampled from a Gaussian distribution with mean and variance, respectively, of
\begin{align}
    \overline{J_{i_1 \dots i_q}} = 0 , \quad \overline{(J_{i_1\dots i_q})^2} = \frac{(q-1)! J^2}{N^{q-1}}.
\end{align}
Our focus is on understanding the properties of systems in which two such SYK models, which we will refer to as SYK ``dots", are coupled by random multi-body interactions \cite{Chen2017,Gu2017a,Haldar2018}, as described by the Hamiltonian,
\begingroup\allowdisplaybreaks
\begin{align}
	H_{q,r} &= \sum_{a=A,B} \sum_{1 \leq i_1 < \dots < i_q\leq N_a} i^{\frac{q}{2}} J^a_{i_1\dots i_q} \chi^a_{i_1} \dots \chi^a_{i_q}  +  \sum_{\substack{1 \leq i_1 <\dots < i_{\frac{r}{2}} \leq N_A \\ 1 \leq j_1 <  \dots < j_{\frac{r}{2}} \leq N_B}} i^{\frac{r}{2}}  V^{i_1 \dots i_{\frac{r}{2}}}_{j_1 \dots j_{\frac{r}{2}}} \chi_{i_1}^A \dots \chi_{i_{\frac{r}{2}}}^A \chi_{j_1}^B \dots \chi_{j_{\frac{r}{2}}}^B \label{eqn:Hqr}
\end{align}\endgroup
where $a=A,B$ labels the two dots. Here, $J^a_{i_1\dots i_q}$ and $V^{i_1 \dots i_{r/2}}_{j_1 \dots j_{r/2}}$ are Gaussian random variables with zero mean and variances,
\begin{align}
	\overline{(J_{i_1\dots i_q}^a)^2} = \frac{(q-1)! J^2}{N_a^{q-1}}, \quad \overline{(V^{i_1 \dots i_{r/2}}_{j_1 \dots j_{r/2}})^2} = \frac{(r/2-1)!^2V^2}{\sqrt{N_A N_B}^{r-1}}.
\end{align}
We assume the variables $J^A_{i_1\dots i_q}$ and $J^B_{i_1\dots i_q}$ are uncorrelated and we take $N_A = N_B \equiv N/2 \in 2 \mathbb{Z}$. For concreteness, we will also fix $q=4$. Our main objects of study will be the $q=r=4$ model, which corresponds to the two-site version of the SYK chain studied in Ref. \cite{Gu2017b}, for which subthermal behavior was observed at large-$N$, and the $q=4$, $r=2$ model, in which the interdot coupling is quadratic in the fermion operators.
In the following, we measure all quantities in units of $J=1$, unless otherwise noted. 

We note that, in the $r=2$ model, the quadratic term is more relevant (in the sense of the renormalization group) than the intradot quartic SYK coupling and hence, at low energies, this model exhibits Fermi liquid like properties \cite{Chen2017}. In particular, while the $r=4$ coupled SYK model exhibits a non-vanishing zero-temperature thermal entropy (like the standard SYK model), the $r=2$ model does not. A detail which will be relevant for 
our finite-$N$ analysis is that while all models under consideration conserve total fermion parity, the $r=4$ model additionally conserves the fermion parity of each SYK dot individually. Moreover, both the $r=4$ model and the original SYK model have an (anti-unitary) particle-hole symmetry, resulting in a doubly degenerate energy spectrum when $N \neq 0 \, \text{mod} \, 8$
\cite{You2017,Cotler2017}. 

\subsection{Quenches}

We will primarily characterize the thermalization of these SYK models by computing the late-time entanglement between the SYK dots $A$ and $B$ after a global quench from a high-energy pure state. 
We recall that given a bipartition of a Hilbert space into subspaces $A$ and $B$ and a density matrix $\rho$, the $n^{\mathrm{th}}$ R\'enyi entropy is given by
\begin{align}
	S^{(n)}_A = \frac{1}{1-n} \log \mathrm{Tr}_{A} \rho^n_A,
\end{align}
where $\rho_A = \mathrm{Tr}_{B} \rho$ is the reduced density matrix for region $A$. The $n \to 1$ limit yields the von Neumann entanglement entropy,
    $S_A = - \Tr  \rho_A \log \rho_A$. 
While we can compute both the R\'enyi and von Neumann entropies in our finite-$N$ ED computations, our large-$N$ numerics only give us access to the R\'enyi entropies; for concreteness, we will focus on $n=2$. In our large-$N$ analysis, we will take as our criterion for thermalization whether the R\'enyi entropies saturate to the value they take in a thermal ensemble with temperature set by the energy of the initial state.
Now, the R\'enyi entropies may not satisfy ETH when we take the subsystem size to be nonvanishing in the thermodynamic limit \cite{Garrison2018}, and so one may argue that it is more appropriate to use the von Neumann entropies as the criteria for thermalization. However, it is still of interest to understand how the R\'enyi entropies thermalize, as they are sensitive to outliers in the entanglement spectrum, and so give more information about the dynamics. We will continue to use them as our criterion for thermalization.

In order to make the R\'enyi entropy calculation tractable in the large-$N$ limit, we choose our initial state such that the R\'enyi entropy can be represented with a simple path integral. 
To that end, we consider quenches from the Kourkoulou-Maldacena (KM) states \cite{Kourkoulou2017}, as in Refs. \cite{Zhang2020b,Liu2020}. We also discuss quenches from thermofield double (TFD) states, as in Refs. \cite{Penington2019,Chen2020,Jian2021}, in Appendix \ref{app:thermofield-double} for comparison. We focus on the KM states in the main text, as the TFD states require a doubling of the Hilbert space, making them less amenable to finite-$N$ analysis.

The KM states are constructed as follows. We first define the infinite temperature KM state by the condition
\begin{align}
	(\chi_{2i-1}^a + i\chi_{2i}^a)\ket{\{1\}} = 0, \quad  a= A,B, \quad \forall \, i. \label{eqn:KM-inf-temp}
\end{align}
This amounts to pairing up the Majorana fermions on each dot into complex fermions, $c_{a,i}^\dagger = \chi_{2i-1}^a + i\chi_{2i}^a$, and setting them to be occupied. It is clear that this state has no entanglement between the $A$ and $B$ subsystems. The finite temperature KM state is then defined through an imaginary time evolution
\begin{align}
	\ket{KM_\beta} = \frac{1}{\sqrt{Z_{KM}}} e^{-\beta H_{q,r} / 2} \ket{\{1\}}, \label{eqn:KM-finite-temp}
\end{align}
where $Z_{KM} = \braket{\{1\}|e^{-\beta H_{q,r}} | \{1\}}$. Note that a non-zero value of $\beta$ introduces entanglement between the $A$ and $B$ fermions. 
After a quench from a KM state, we expect equilibration to a steady-state at effective inverse temperature on the same order as $\beta$. In particular, at least in large-$N$, the late-time R\'enyi entropies should approach those of the canonical ensemble described by $\rho_\beta = e^{-\beta H_{q,r}}/Z(\beta)$, with $Z(\beta) = \mathrm{Tr}[e^{-\beta H_{q,r}}]$.

The KM states are states of definite fermion parity of each SYK dot and hence total fermion parity as well.
Quenches from these states will therefore only involve half the Hilbert space in the $r=2$ model and one quarter of the Hilbert space in the $r=4$ model. 
Although this is unimportant in the large-$N$ limit, it will prove helpful in our finite-$N$ analysis to consider quenches from states mixing all fermion parity sectors when comparing these models. 
To that end, we define a new state $\ket{\widetilde{KM}_\beta}$ 
\begin{align}
    \begin{split}
    \ket{\widetilde{KM}_\beta} = \frac{1}{2} & e^{-\beta H_{q,r} /2} \left[ \left( \ket{0 1 \dots 1}_A + \ket{1 1 \dots 1}_A  \right) \otimes \left( \ket{0 1 \dots 1}_B + \ket{1 1 \dots 1}_B  \right) \right]. \end{split} \label{eqn:KM-indefinite-parity}
\end{align}
Here, $\ket{n_1 , n_2, \dots, n_{N/4}}_a$ is a state with occupation number $n_i$ for the fermion $c_{a,i}$ on dot/subsystem $a=A,B$. This is simply a superposition of KM-like states which is neither an eigenstate of the total fermion parity nor the fermion parity in each subsystem. As a consequence, time evolution after a quench from this state will involve the full Hilbert space in both coupled models.

Operationally, our numerical quench experiments then proceed as follows. Starting with either the $r=2$ or $r=4$ Hamiltonian, we first fix the disorder realization and then time evolve the KM state with a given $\beta$ for a sufficiently long time, which will still be $O(1)$ relative to $N$ in the large-$N$ limit, until the system reaches equilibrium. Here, by equilibrium, we mean that the macroscopic quantities of interest, like the R\'enyi entanglement entropies, do not change appreciably at later times. We then repeat the quench experiment with a different disorder realization and a KM state with the same $\beta$.\footnote{Note that since the KM state for $\beta \neq 0$ depends on the Hamiltonian, the initial state we quench from also varies between disorder realizations.} We then average the late-time values of the quantities of interest, including the R\'enyi entropy, over all disorder realizations.

\section{State-Dependent Thermalization in Large-N Quenches \label{sec:large-N} }

Our first goal is to 
characterize the thermalization of the coupled SYK models after quenches from the 
KM states in the large-$N$ limit. As noted in the Introduction, the SYK model is exactly solvable in this limit, a feature which can been exploited to numerically compute the R\'enyi entropies 
\cite{Penington2019,Chen2020,Zhang2020a,Haldar2020,Zhang2020b,Jian2021}.
In this section, we first briefly review the path integral setup of the R\'enyi entropies for the 
KM state quenches, after which we present our numerical results. 
We will find that the $r=4$ model \emph{can} thermalize, provided the effective temperature of the initial state is high enough. 
This is in contrast to the $r=2$ model, which appears to always thermalize. As we will review, the process of thermalization in the large-$N$ limit is governed by an interplay between so-called replica diagonal and non-diagonal solutions to the saddle-point equations \cite{Penington2019,Chen2020}. We provide additional data for thermofield double (TFD) state quenches in Appendix \ref{app:thermofield-double}.

\subsection{Path Integral Setup of KM Quench}


\begin{figure}
  \includegraphics[width=0.9\textwidth]{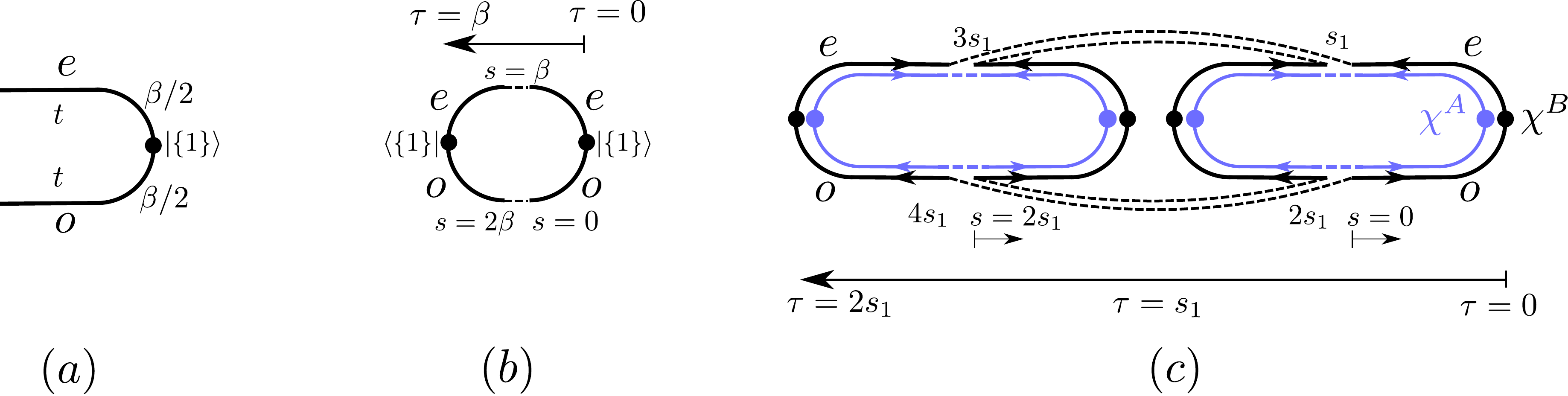}
  \caption{ (a) Pictorial representation of $\ket{KM_\beta}$. The dot represents the boundary condition of Eq. \eqref{eqn:KM-inf-temp} and the two branches labelled by $e$ and $o$ the evolution of the $\chi_{2i-1}^a$ and $\chi_{2i}^a$ fermions, respectively. The curved (horizontal) lines represent imaginary (real) time evolution by $\beta/2$ ($t$).  (b) Contour $\mathcal{C}$ used in the evaluation of $Z_{KM}$. Here, $\tau$ parameterizes the imaginary time evolution by $\beta$ of the $e$ and $o$ fermions. The contour can be reparameterized by new fermions $\tchi^a_i(s)$, where $s$ increases counter-clockwise around the contour. (c) Contour $\mathcal{C}'$ used in the evaluation of $S_A^{(2)}$, where $\tau$ paramterizes the evolution of the $e$ and $o$ fermions. The dashed lines indicate the boundary conditions of the fermions. The variable $s$ parameterizes the contour and increases counter-clockwise around each replica. Here, $s_1 = 2t+\beta$. 
  } 
  \label{fig:km-setup}
\end{figure}

The second R\'enyi entropy after a quench from the KM state can be evaluated in terms of a path integral defined on an appropriate contour. As this setup has been discussed previously (see e.g. Refs. \cite{Zhang2020b,Liu2018}), we outline the construction here and relegate the details to Appendix \ref{app:discretization}. 

As a step towards computing the R\'enyi entropy, let us first find a path integral expression for the normalization $Z_{KM}$ in Eq.~\eqref{eqn:KM-finite-temp}. We can pictorially represent the time evolved KM state $\frac{1}{Z_{KM}} e^{-(it+\beta/2) H_{q,r}} \ket{\{1\}}$ as in Fig. \ref{fig:km-setup}(a). Here, we represent the imaginary and real time evolution of the even ($e$) and odd ($o$) fermions $\chi_{2i}^a$ and $\chi_{2i-1}^a$, respectively with separate legs emanating from the state $\ket{\{1\}}$, represented by the dot. 
The normalization $Z_{KM}$ can then be expressed as a path integral
\begin{align}
	Z_{KM} = \braket{ \{1\} | e^{-\beta H_{q,r}} | \{1\} } = \int_{\mathrm{b.c.}} \mathcal{D}\chi_i^A \chi_i^B e^{- I_{\mathcal{C}} }; \quad
	I_{\mathcal{C}} = \int_{\mathcal{C}} d\tau \left[ \frac{1}{2}\sum_i \chi_i \partial_\tau \chi_i + f(\tau) H_{q,r}\right], \label{eqn:km-action-unaveraged}
\end{align}
where $I_{\mathcal{C}}$ is evaluated on the contour $\mathcal{C}$ shown in Fig. \ref{fig:km-setup}(b), parameterized by $\tau \in [0,\beta)$ and obtained by taking two copies of Fig. \ref{fig:km-setup}(a) and connecting the matching legs.
The function $f(\tau)$ accounts for whether we are performing imaginary time ($f(\tau) = 1$), forward real time ($f(\tau)=i$), or backward real time ($f(\tau) = -i$) evolution at the point $\tau$ on the contour. 
There is no real time evolution involved in computing the normalization, so we have $f(\tau) \equiv 1$ for this contour.
The ``b.c." in the integral limit indicates that the fields are subject to the boundary conditions
\begin{align}
    \begin{split}
	\chi_{2j-1}(\tau = 0) &= -i \chi_{2j}(\tau =  0), \quad	\chi_{2j-1}(\tau =  \beta) = i\chi_{2j}(\tau =  \beta),
	\end{split} \label{eqn:km-bcs}
\end{align}
imposed by the condition of Eq. \eqref{eqn:KM-inf-temp}. 
Now, as the pictorial representation of the contour suggests, and as we review in Appendix \ref{app:discretization}, we can re-express the path integral as one over $N/2$ Majorana fermions $\tchi_i^a(s)$ defined on $s \in [0,2\beta)$; the reparameterization of the contour in terms of $s$ is depicted in Fig. \ref{fig:km-setup}(b).

Our interest is in the disorder averaged quantity $-\overline{\log Z_{KM}}$, where the overline indicates the average over all realizations of the couplings. Although the disorder average can be handled with a replica trick, we will make the standard assumption of a disorder replica diagonal solution in the treatment of SYK models, allowing
us to approximate $-\overline{\log Z_{KM}} \approx -\log \overline{Z_{KM}}$. 
We can thus directly average the path integral over all disorder realizations. Following standard manipulations \cite{Maldacena2016}, we can introduce the Lagrange multiplier fields $\Sigma_a(s,s')$ to enforce the constraint
\begin{align}
    G_a(s,s') &= \frac{1}{N_a}\sum_i \tchi_i^a(s) \tchi_i^a(s').
\end{align}
On-shell, $\Sigma_a(s,s')$ and $G_a(s,s')$ represent the self-energies and Green functions, respectively, of the $\tchi^a$ Majorana fermions. Note that we have introduced separate Green functions for the $a=A,B$ fermions, as when we next consider the R\'enyi entropy computation, they will obey distinct boundary conditions in $s$. Now, upon disorder averaging and integrating out the fermions in favor of the above bilocal fields, we obtain the effective action (we will henceforth drop overlines as disorder averages will always be assumed)
\begin{align}
    \begin{split}
	\frac{I_{\mathcal{C}}}{N} =  &-\frac{1}{8} \log \det \left(  \underset{a}{\partial_s} - \Sigma_{a}  \right) + \frac{1}{8}\int_{\mathcal{C}} ds ds'  \Bigg[ \sum_{a} \Sigma_{a}(s,s') G_{a}(s,s') \\
	&\qquad\qquad - F(s,s')P(s,s') \Bigg( \frac{J^2}{q} \sum_{a} G_{a}(s,s')^q  + \frac{2V^2}{r} G_A(s,s')^{\frac{r}{2}} G_B(s,s')^{\frac{r}{2}} \Bigg) \Bigg] \, ,\end{split}
	\label{eqn:km-action}
\end{align}
where $F(s,s')=f(s)f(s')$. The subscript $a$ on $\partial_s$ is included to keep track of the temporal boundary conditions for the $\tchi^A$ and $\tchi^B$ fermions. In computing the normalization, both obey the same, standard fermionic boundary conditions. 
In deriving the effective action, it is necessary to introduce the function $P(s,s')$, which is defined as
\begin{align}
    P(s,s') = \begin{cases}
    1 & s,s' \text{ both on an }e \text{ or } o \text{ contour leg} \\
    0 & \text{otherwise}.
    \end{cases}
\end{align}
The corresponding saddle-point or Schwinger-Dyson, equations are given by,
\begin{align}
    \begin{split}
	G_a &= (\underset{a}{\partial_s} - \Sigma_{a})^{-1} , \quad
	\Sigma_a(s,s') = P(s,s')F(s,s') \Big[J^2 G_a^{q-1}(s,s') + V^2 G_a^{r/2-1}(s,s') G_{\bar{a}}^{r/2}(s,s')\Big] , \label{eqn:km-sd-eqns}
	\end{split}
\end{align}
where if $a=A$, then $\bar{a}=B$ and vice versa.
The Schwinger-Dyson equations can be discretized and solved numerically via a standard self-consistent iterative procedure, the details of which are provided in Appendix \ref{app:discretization}. 
In the large-$N$ limit, the path integral is dominated by this saddle-point contribution, and so evaluating Eq.~\eqref{eqn:km-action} on-shell thus gives us the normalization, $Z_{KM}$, of the KM state.

The computation of the R\'enyi entropy after a KM state quench proceeds analogously. We must evaluate 
\begin{align}
    e^{-S^{(2)}_A} = \Tr_A [( \Tr_B \ket{KM_\beta}\bra{KM_\beta})^2 ]=
	\frac{1}{Z_{KM}^2} \int_{\mathrm{b.c.}} \mathcal{D}\chi_i^A \chi_i^B e^{- I_{\mathcal{C}'} }, \label{eqn:km-trace}
\end{align}
where we have again used a path integral representation. Here $I_{\mathcal{C'}}$ takes the form of Eq. \eqref{eqn:km-action-unaveraged}, evaluated on the contour $\mathcal{C}'$, shown in Fig. \ref{fig:km-setup}(c) and parameterized by $\tau \in [0,2s_1)$, where we set $s_1 = 2t+\beta$. We arrive at the contour $\mathcal{C}'$ by taking two replicas of the KM density matrix, formed from Fig. \ref{fig:km-setup}(a). The boundary conditions for the $\chi^{a}_i$ fermions, indicated by the dashed lines in Fig. \ref{fig:km-setup}(c), follow from the partial traces in the above expression for the R\'enyi entropy. Again by introducing new fermions and reparameterizing the contour by $s \in [0,4s_1)$ (see Appendix \ref{app:discretization}) we find that the disorder-averaged action -- assuming once more a disorder replica diagonal solution -- takes the form of Eq. \eqref{eqn:km-action} evaluated on the contour $\mathcal{C}'$  and with both $F(s,s')$ and $P(s,s')$ appropriately modified. 
In particular, we indicate where $f(s) = +1, +i, -i$ on the contour by, respectively, curved segments, horizontal segments with a leftward pointing arrow, and horizontal segments with a rightward pointing arrow.
After numerically solving the saddle-point equations, we finally compute the second R\'enyi entropy as
\begin{align}
    S^{(2)}_A(t) = - I_{\mathcal{C}'} + 2I_{\mathcal{C}},
\end{align}
where the actions $I_{\mathcal{C}'}$ and $I_{\mathcal{C}}$ are evaluated on-shell. 

\subsection{Numerical Results}

We now present our numerical results for the post-quench R\'enyi entropies. We focus on the KM states here 
and provide additional discussion for TFD state quenches in Appendix \ref{app:thermofield-double}. 
While we reproduce some prior results in the literature (e.g. the same numerical computation was considered for $r=4$ and a TFD quench in Ref. \cite{Penington2019}), our main observation which, to our knowledge, has not been previously reported is the state-dependent thermalization of the R\'enyi entropies in the $r=4$ model for certain parameter ranges. 


\begin{figure}
\subfloat[$r=4,V=0.35J$]{%
    \includegraphics[width=0.33\linewidth]{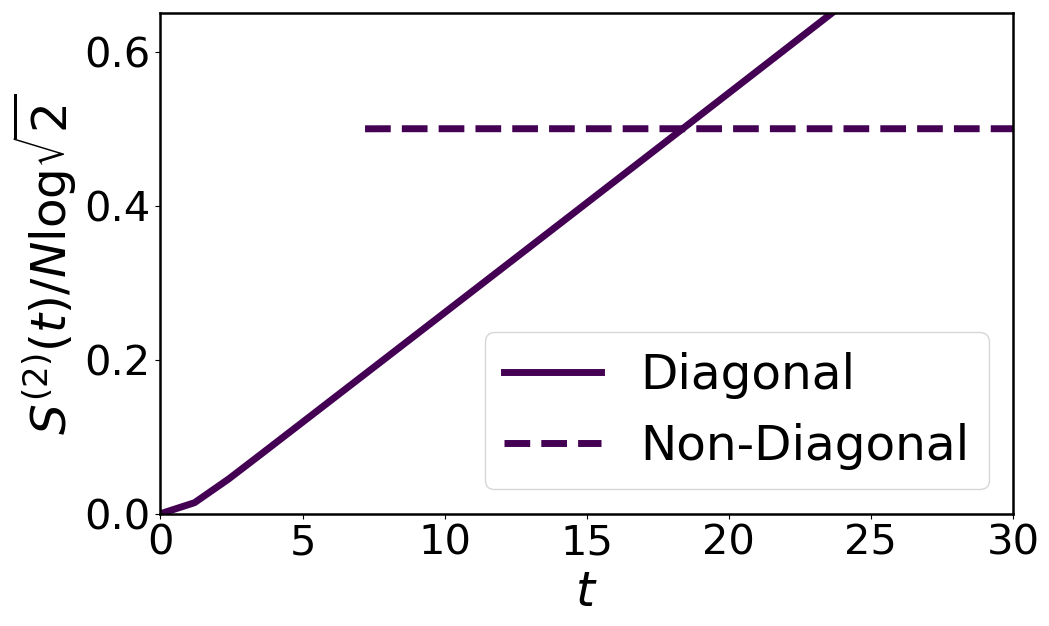}
} \hspace{15mm}
  \subfloat[$r=2, V=0.1J$]{%
    \includegraphics[width=0.33\linewidth]{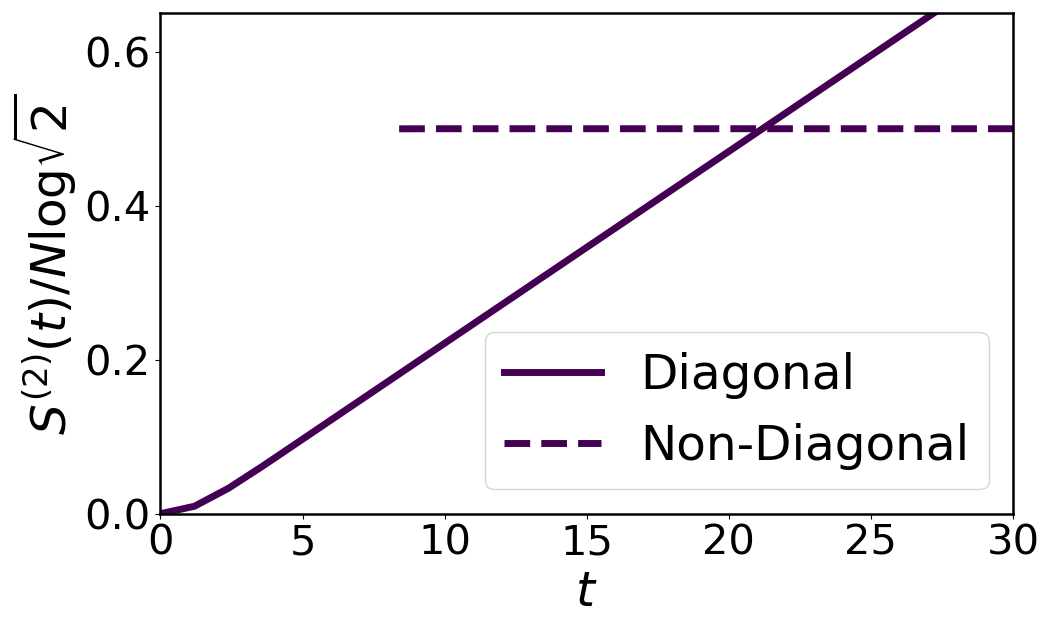}
}
  \caption{Time dependence of the second R\'enyi for the (a) $r=4$ with $V=0.35J$ and (b) $r=2$ with $V=0.1J$ models after a quench from the $\beta=0$ KM state. The solid (dashed) lines represent the replica diagonal (non-diagonal) saddle-points. 
  } 
  \label{fig:two-site-beta0}
\end{figure}

We first consider a quench from the $\beta=0$ KM state in both the $r=2$ and $r=4$ models, taking a moderate value of $V < J$. As shown in Fig. \ref{fig:two-site-beta0}, there are two solutions to the saddle-point equations. The first is the replica-diagonal saddle-point \cite{Penington2019,Chen2020}, so named as it does not involve correlations between the replicas involved in the R\'enyi entropy calculation (i.e. $G_a(s,s')$ vanishes for $s$ and $s'$ belonging to disconnected $\chi^a$ contours in Fig. \ref{fig:km-setup} (c)). 
This saddle-point yields the linearly increasing entropy indicated by the solid lines. The second saddle-point is replica non-diagonal, involving correlations between the replicas, and appears as a solution above a critical value of $t$. This saddle-point has a nearly $t$-independent action, indicated by the horizontal dashed lines. 
In particular, the non-diagonal saddle-point yields the maximal  R\'enyi entropy of $S^{(2)} = (N/2)\log\sqrt{2}$ for an equal bipartitioning of a system of $N$ Majorana fermions, and so describes an equilibrium state at infinite temperature, as expected for a quench from a $\beta=0$ KM state. 
The physical value of the entropy after the quench is given by the minimum entropy of the two saddle-points. So we see that in both models, after a period of linear growth governed by the replica diagonal saddle-point, the system switches in $O(1)$ time to the replica non-diagonal saddle-point, at which point the system has thermalized. 
Hence the R\'enyi entropies in the $r=4$ model can thermalize, at least for effectively infinite temperature states.\footnote{We note that similar behavior in effective infinite temperature quenches has been observed in related models, such as the Brownian SYK model \cite{Jian2021} and non-unitary SYK chain (including when non-Hermitian couplings are turned off) \cite{Liu2020}. This behavior for the $r=4$ model was also discussed in a quench in the microcanonical ensemble in Ref. \cite{Penington2019}. }

\begin{figure}
\subfloat[$V=0.35$, $J=1$]{%
    \includegraphics[width=0.45\linewidth]{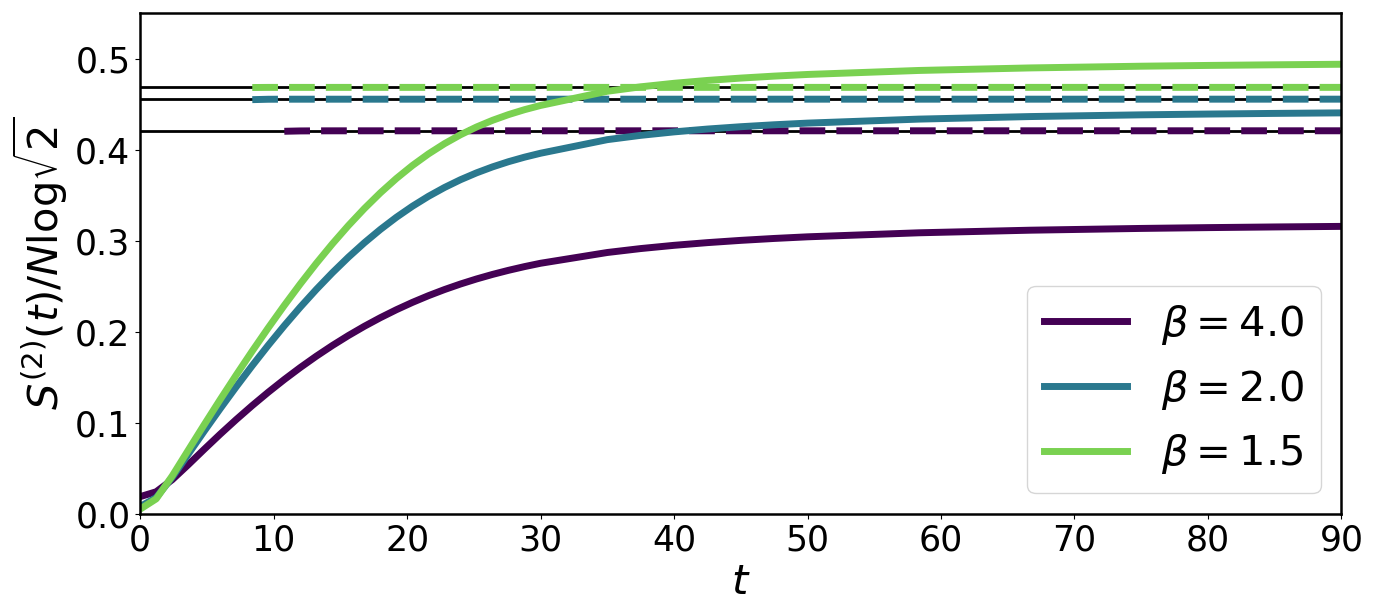}
}
\subfloat[$V=0.35$, $J=0$]{%
    \includegraphics[width=0.45\linewidth]{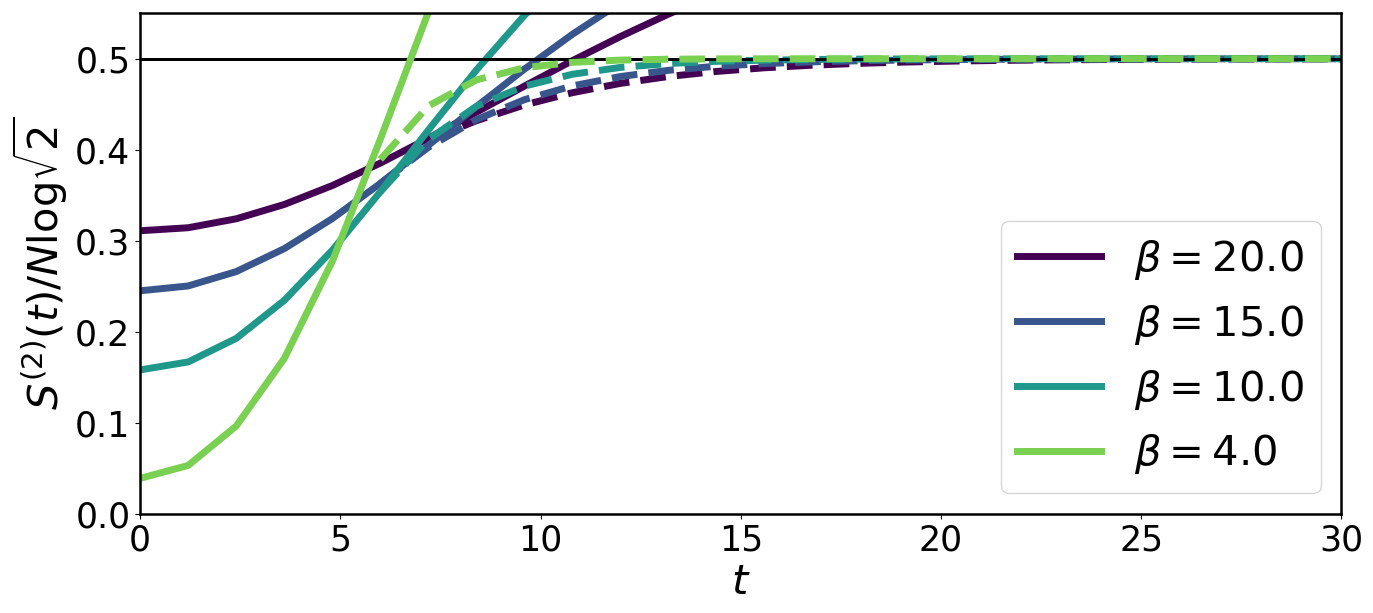}
}
  \caption{ Second R\'enyi after a quench from the KM state with different $\beta$ in the $r=4$ model for $V=0.35$ and (a) $J=1$ (b) $J=0$. The solid (dashed) lines represent the replica diagonal (non-diagonal) saddle-points. 
  The black lines indicate the R\'enyi entropy in the canonical ensemble for each $\beta$. Note that in (b), the thermal data for different $\beta$ are too close to distinguish. 
  } \label{fig:r4-km-largeN}
\end{figure}

Quenches from KM states for $r=4$ with non-zero $\beta$ exhibit qualitatively distinct behavior, as shown in Fig. \ref{fig:r4-km-largeN}(a), in which we plot the R\'enyi entropy for $V=0.35$ and different values of $\beta$. 
Notice that the entropy for the non-diagonal saddle-point is now shifted down from the maximal value due to finite $\beta$. In this case we shall instead compute the R\'enyi entropy in the canonical thermal ensemble at the same effective temperature $\beta$ -- which is the ensemble we expect to describe the late-time behavior -- using the same numerical techniques \cite{Zhang2020a} (see Appendix \ref{app:discretization}). 
These values are indicated by the horizontal black lines in Fig. \ref{fig:r4-km-largeN}, from which we see the non-diagonal solution yields an entropy corresponding exactly to the expected thermal value, and hence still describes a fully thermalized state. However, as shown in Fig. \ref{fig:r4-km-largeN}(a), for $\beta \gtrsim  2$, the diagonal solution plateaus to a value below that of the non-diagonal solution. 
That is to say, the system does not equilibrate to the expected effective temperature, and so we thus reproduce the \emph{subthermal} behavior of the R\'enyi entropy reported in earlier works.
On the other hand, we also see that for $\beta \lesssim 2$, the diagonal saddle-point crosses the non-diagonal saddle-point in $O(1)$ time. For this and smaller values of $\beta$, the system is thus in a \emph{thermal} state at late times. If we completely turn off the intradot couplings (i.e. set $J=0$) then, as shown in Fig. \ref{fig:r4-km-largeN}(b), the system thermalizes for all $\beta$ considered. Though we cannot say for certain, following the trend for increasing $\beta$ in this case suggests that the system will thermalize for all values of $\beta$.

What we have thus illustrated is that the $r=4$ model in fact exhibits \emph{state-dependent} thermalization: for a moderate value of $V/J$, the R\'enyi entropy reaches a thermal (subthermal) value after a quench from a KM state with effective temperature below (above) some critical value $\beta_c$.
In Appendix \ref{app:thermofield-double}, we point out that the same is true for TFD quenches. In fact, in that case, the system never thermalizes for sufficiently small values of $V$ and any $\beta \neq 0$, while for $J=0$ there remains a range of $\beta$ for which the system still exhibits a subthermal R\'enyi entropy.
Although increasing $\beta$ in the KM and TFD states introduces additional entanglement between $A$ and $B$ in the initial state, it also suppresses the contribution from higher-energy states, indicating that the $r=4$ model exhibits different behavior at different energy scales. 
In particular, in the limit $\beta \to \infty$, both $\ket{KM_\beta}$ and $\rho_\beta = e^{-\beta (H)}/Z(\beta)$ should describe the ground state (provided it has non-zero overlap with $\ket{\{1\}}$). However, we see from Fig. \ref{fig:r4-km-largeN}(a) that the entanglement of $\ket{KM_\beta}$ must saturate to a value below that of the ground state value as $\beta$ increases; 
this is suggestive of the possibility that, in the large-$N$ limit, some degrees of freedom are ``frozen" and do not contribute to the entanglement.

This appears to be consistent with Ref. \cite{Gu2017b}, which argued that the observed subthermal behavior may be a consequence of ``heavy" modes becoming localized in the large-$N$ limit. 
These heavy modes were also identified as contributing to the zero-temperature thermal entropy of the $r=4$ model. One may then expect that a model with vanishing zero-temperature thermal entropy in the large-$N$ limit would not support such frozen heavy modes and hence would always thermalize. 

\begin{figure}
\subfloat[$r=2, \, V=0.35J$]{%
    \includegraphics[width=0.45\linewidth]{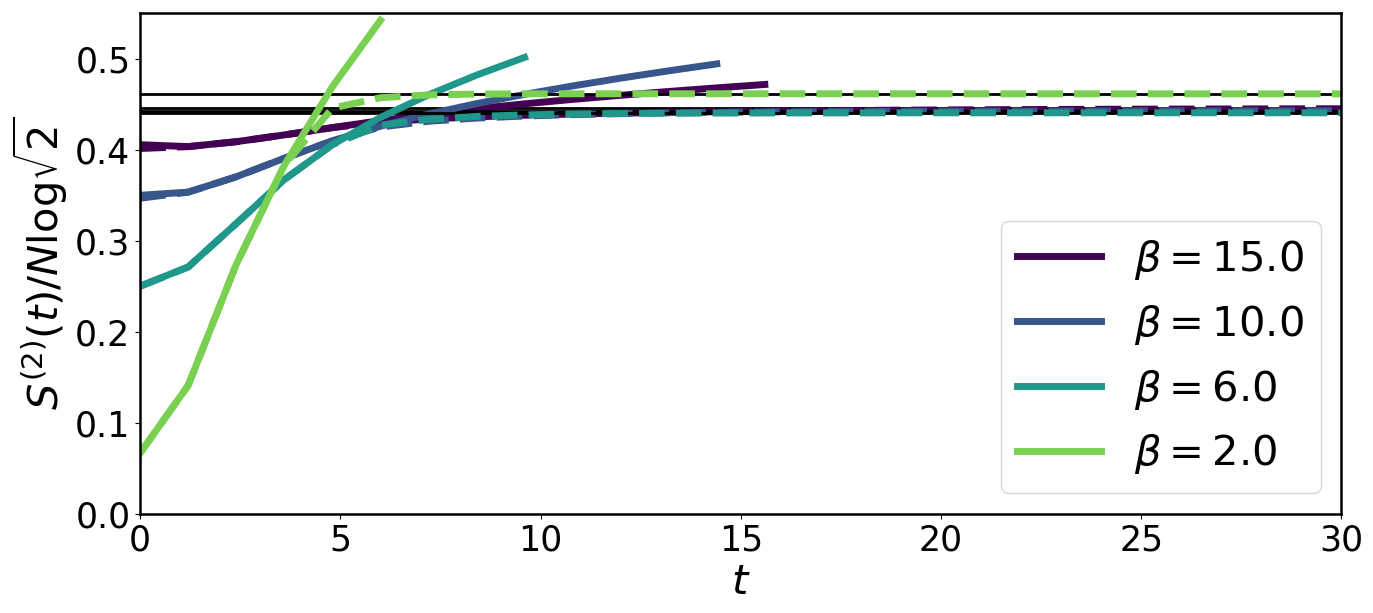}
}
\subfloat[$r=2, \, V=0.1J$]{%
    \includegraphics[width=0.45\linewidth]{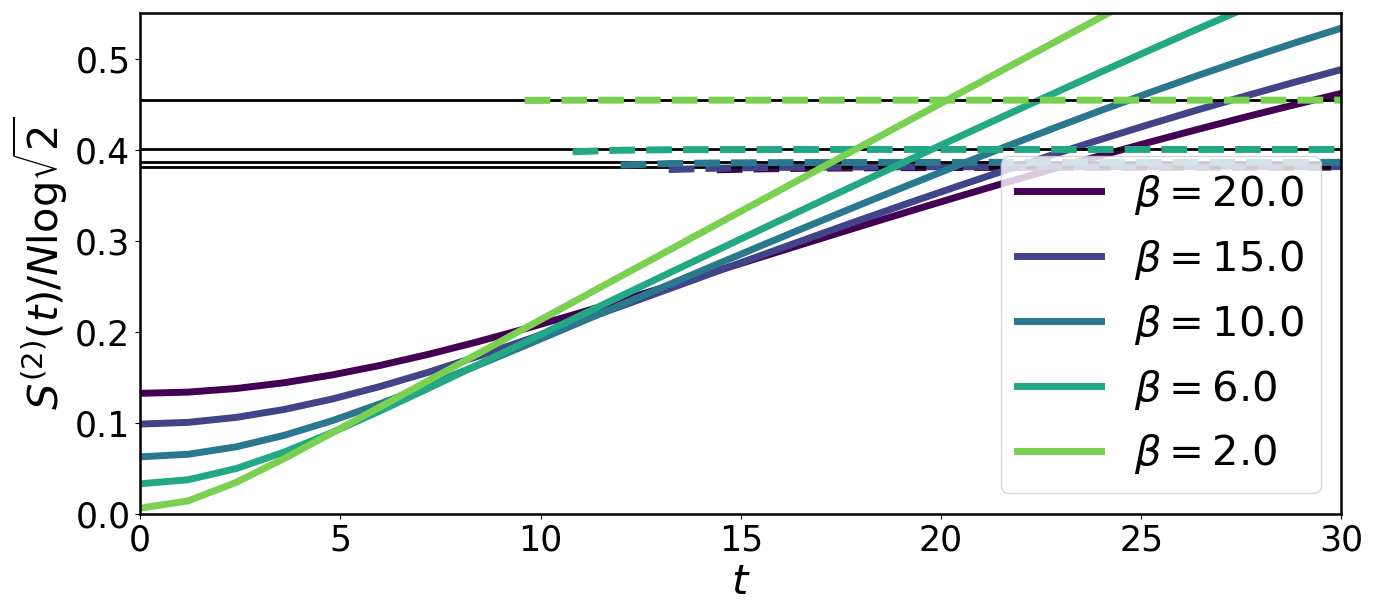}
}
  \caption{ Second R\'enyi after a quench from the KM state in the $r=2$ model with different $\beta$, $J=1$, and (a) $V=0.35J$ (b) $V=0.1J$. The plots follows the same labelling conventions as in Fig. \ref{fig:r4-km-largeN}. 
  } \label{fig:r2-km-largeN}
\end{figure}

As noted above, the $r=2$ coupled model is precisely such a system. 
With this in mind, we turn to finite-$\beta$ quenches in this model, the data for which is plotted in Fig. \ref{fig:r2-km-largeN}. We again find both replica diagonal and non-diagonal saddle-points, and the system always appears to reach the diagonal solution in finite time and thus thermalizes.
Of course, we cannot rule out the possibility that the $r=2$ model does not thermalize for sufficiently large $\beta$ and small $V$.
However, as we see from Fig. \ref{fig:r2-km-largeN}, increasing $\beta$ appears to increase the initial R\'enyi entropy faster than the late-time non-diagonal value decreases, suggesting that even if the replica-diagonal saddle-point saturates at late-time, it will likely saturate to a value above the non-diagonal solution, even as $\beta \to \infty$. 
Again, we cannot say this with certainty, but we will continue under the assumption that the $r=2$ model does indeed always thermalize.

Before proceeding, we note that as claimed in Ref. \cite{Penington2019} in which the $r=4$ model was studied at large-$N$, it is possible that the entanglement entropy may thermalize even if the R\'enyi entropies do not, if there are indeed an extensive number of non-thermalizing ``heavy" modes. This is a consequence of the sensitivity of the R\'enyis to tails in the entanglement spectrum induced by the presence of these non-thermalizing modes. Additionally, it was noted in Ref. \cite{Penington2019} that the R\'enyi entropies in the $r=4$ model can thermalize after a quench in the microcanonical ensemble. Although one may thus object to describing the system as ``subthermal" based on the equilibrium values of the R\'enyi entropies, we will continue in our use of this terminology, as our interest lies in part in the fact that the R\'enyi entropies are sensitive to outliers in the entanglement spectrum and in providing further evidence that indeed there are non-thermalizing heavy-modes.

\section{Finite-N Spectral Properties \label{sec:spectral-properties} }

Having illustrated the state-dependent thermalization of the $r=4$ model in the large-$N$ limit, we now wish to see whether vestiges of this behavior also appear in finite $N$ systems. To this end, in the balance of this manuscript, we investigate the coupled SYK models using exact diagonalization (ED). For concreteness, we fix $V=0.35$ in both the $r=2$ and $r=4$ coupled models here and in all subsequent analysis.
Before reexamining the entanglement dynamics after quenches in these models in Section \ref{sec:finite-N-quenches}, we first make a brief detour in the present section and discuss their spectral properties to build some intuition as to whether or not they should thermalize.

\subsection{Spectral Form Factor}

To begin, we consider the spectral form factor (SFF), defined as
\begin{align}
    K(\tau) = \langle \sum_{i,j} e^{i(E_i - E_j) \tau} \rangle,
\end{align}
which has become a vital tool in the study of quantum many-body chaos \cite{Cotler2017,Liu2018sff}.
Here, $E_i$ are the energies, $\tau$ is a fictitious time, and the angular brackets indicate a disorder average. This quantity captures correlations between energy levels, with the distance between the energy levels being probed parameterized by $\tau^{-1}$.
The expectation is that ``chaotic" systems should have a SFF similar to that of a random matrix ensemble with the appropriate symmetries. In particular, RMT predicts that the SFF first decreases as a function of $\tau$, a feature known as the ``slope", then increases linearly -- the ``ramp" -- as a function of $\tau$, and then finally saturates to a ``plateau". The spectral rigidity 
characteristic of random Hamiltonians is what leads to the ramp.
We may also compute the connected part of the SFF,
\begin{align}
    K_c(\tau) = K(\tau) - |\langle\sum_{i} e^{iE_i \tau} \rangle|^2,
\end{align}
which subtracts out non-universal correlations at small $\tau$ in the slope.  

\begin{figure}
\centering
\subfloat[]{%
  \includegraphics[width=0.33\columnwidth]{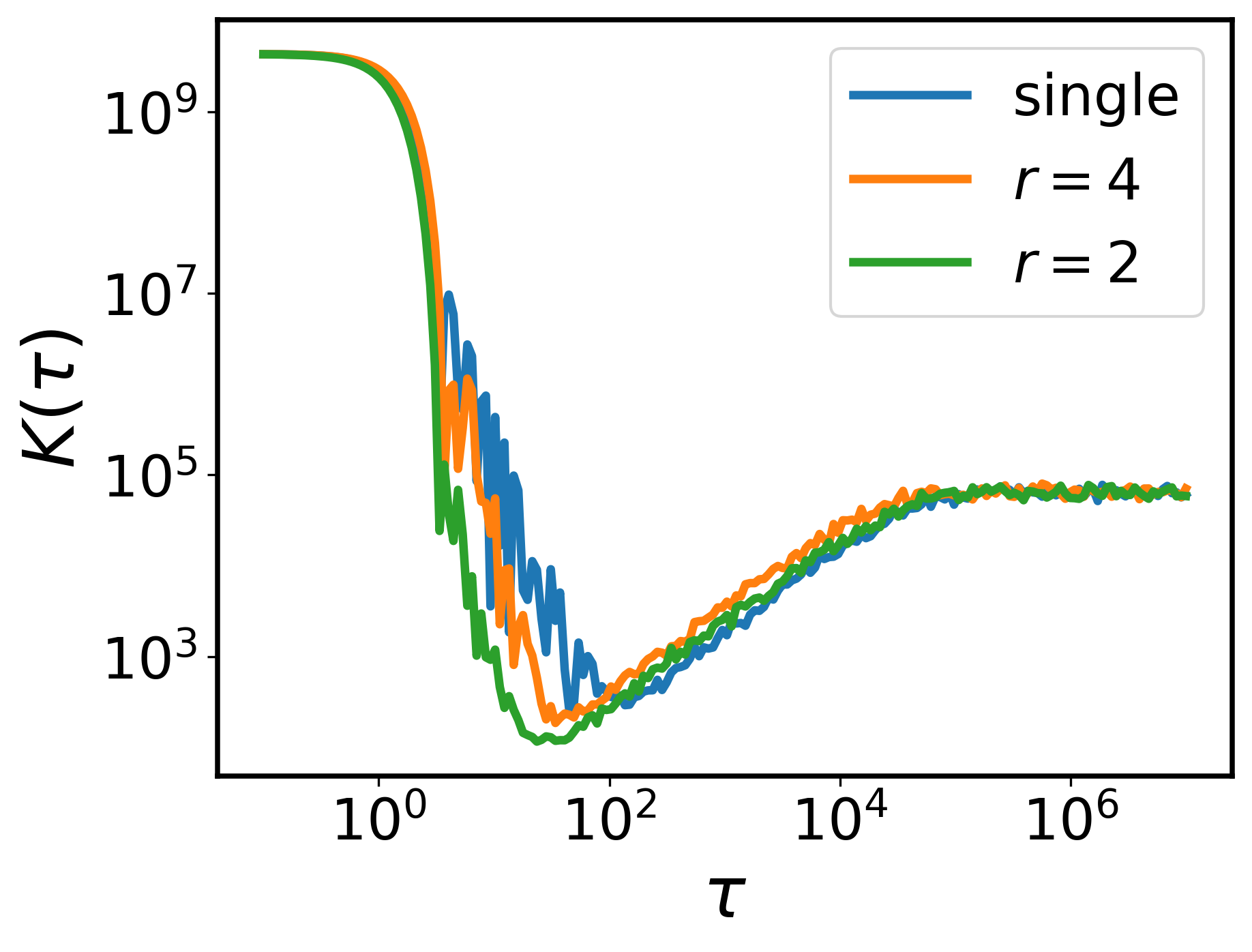}
}\hspace{15mm}
\subfloat[]{%
  \includegraphics[width=0.33\columnwidth]{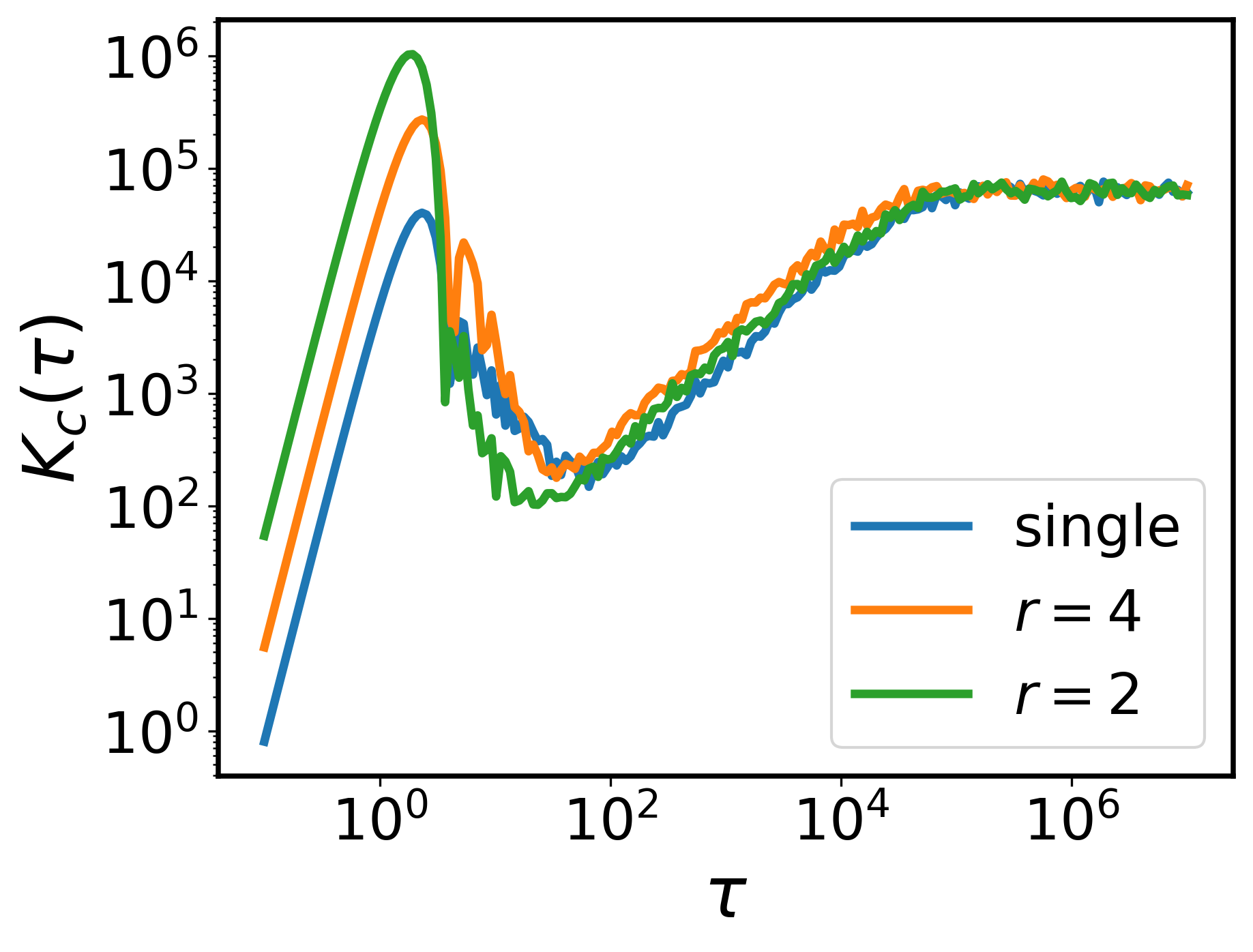}
}
  \caption{(a) Spectral form factors and (b) connected spectral form factors for the $r=2$ and $r=4$ models with $V=0.35J$ compared against the single SYK model for $N=32$, averaged over $100$ disorder realizations. 
  \label{fig:syk-sff} }
\end{figure}

Now, if the the large-$N$ subthermal behavior of the $r=4$ model is a consequence of localized or conserved quantities, this should manifest in weakened energy level repulsion and hence in a deviation of the SFF from RMT predictions.
To that end, we present numerical computations of the SFF for the coupled models compared against that of the single SYK model in Fig. \ref{fig:syk-sff}. 
We take $N=32$ so that all three models fall in the same RMT symmetry class.
We see that the SFFs of both coupled models are qualitatively similar to that of the single SYK model, exhibiting the characteristic slope-ramp-plateau structure of SFFs for Gaussian random matrices. While the $r=2$ ramp agrees quite well with the single SYK ramp, the $r=4$ ramp appears linear albeit with a slightly larger slope. 
Nevertheless, the persistence of the ramp over the same range of $\tau$ as the single SYK implies the presence of energy level anti-correlation, albeit perhaps slightly weakened, over the same energy scales.

Additionally, all three models exhibit an ``overshoot" of the ramp from the slope in the connected SFF, which is more pronounced in the coupled SYK models. 
These deviations from RMT are perhaps to be expected, as the coupled SYK Hamiltonians are necessarily more sparse than that of the single SYK, being more local and hence involving fewer couplings, and so should resemble less a truly random matrix. In spite of this enhanced locality, we see that the coupled SYK models are still clearly quantum chaotic by this measure of spectral rigidity. Hence, any potentially localized heavy modes present in the large-$N$ limit, which may lie at the origin of subthermal behavior in that limit, likely do not manifest at finite values of $N$.

\subsection{Eigenstate Entanglement}

\begin{figure}
  \subfloat[$r=4$]{%
  \includegraphics[width=0.25\textwidth]{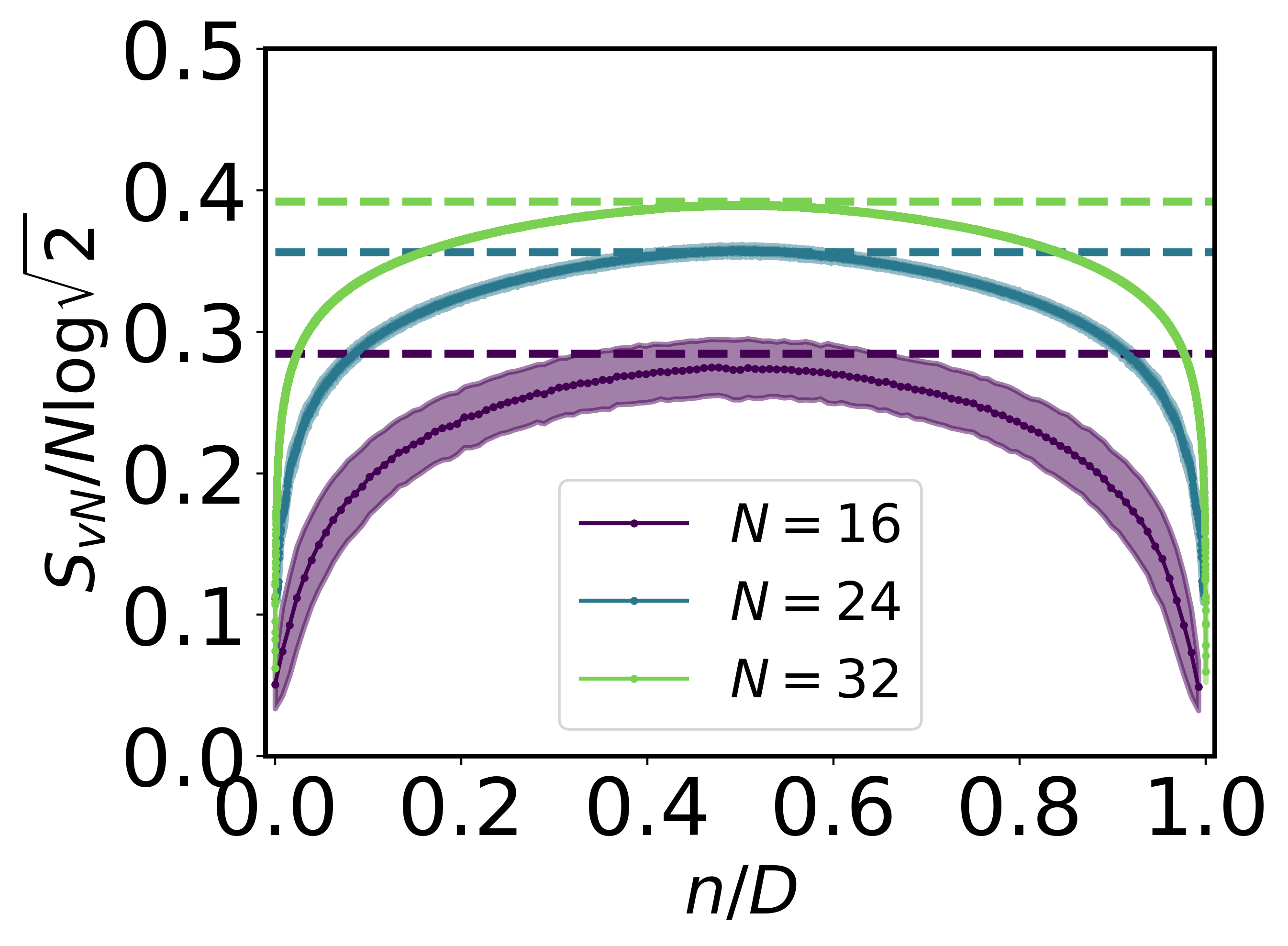}%
  \includegraphics[width=0.25\textwidth]{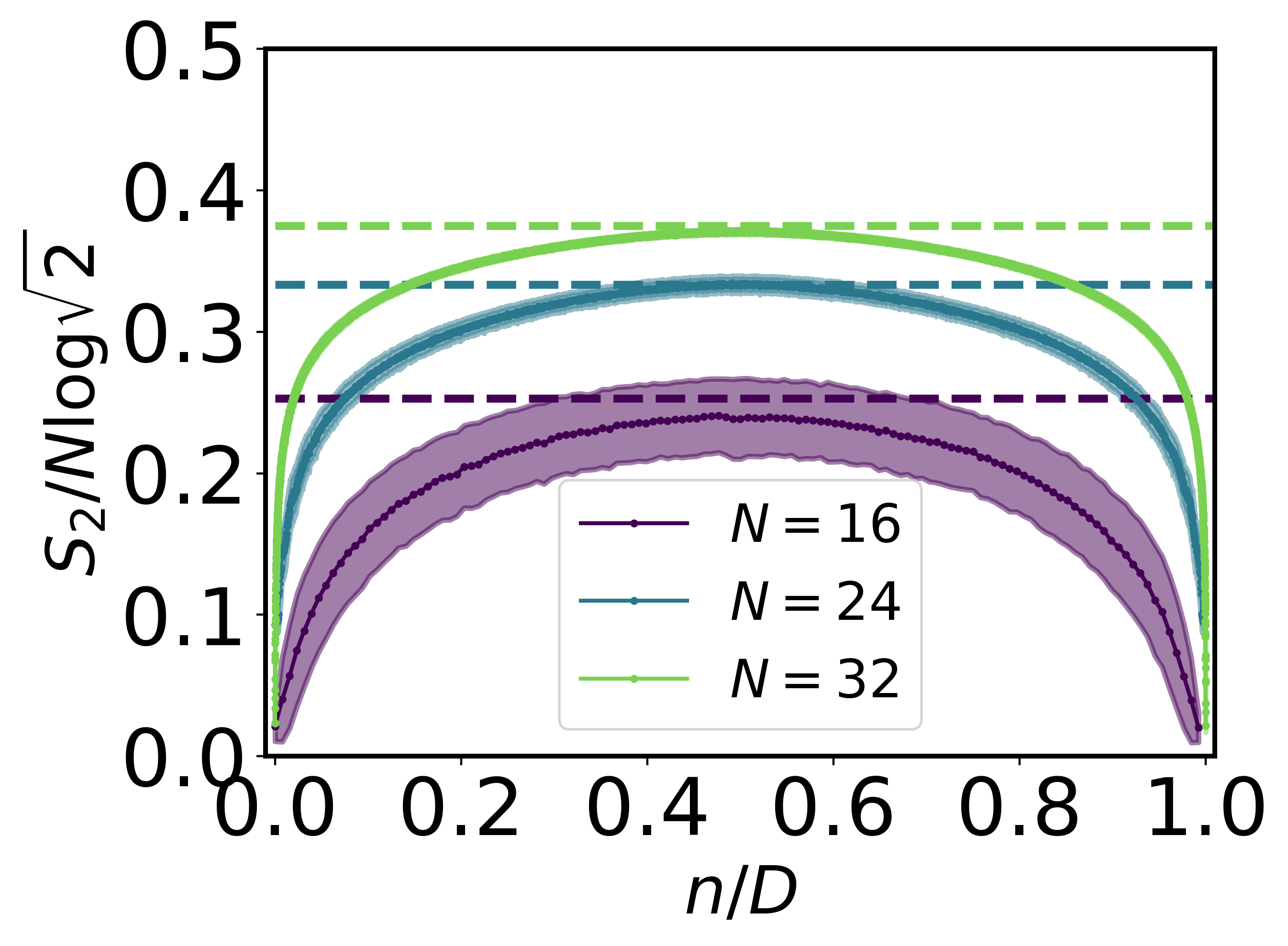}
  }
  \subfloat[$r=2$]{%
  \includegraphics[width=0.25\textwidth]{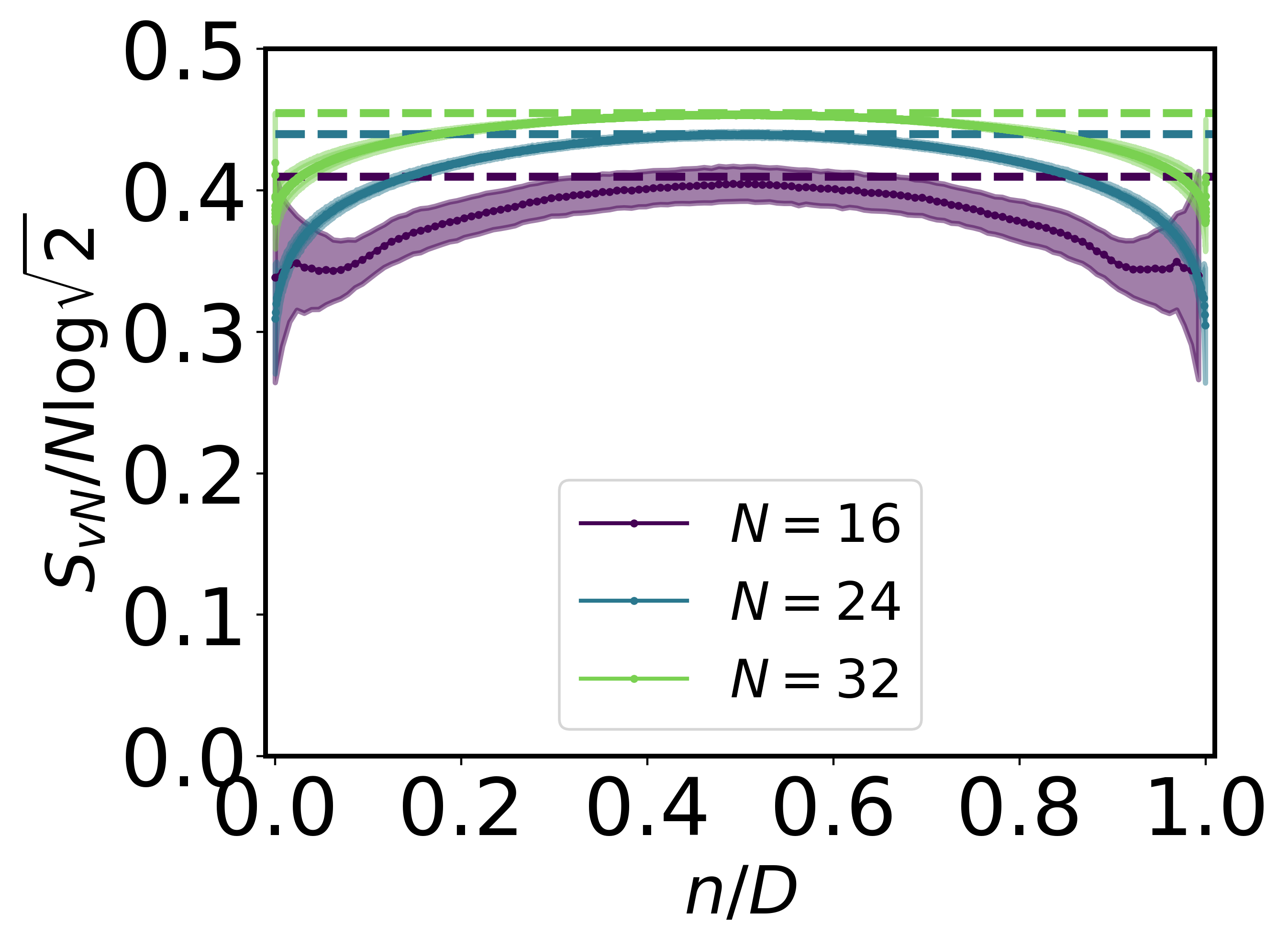}%
  \includegraphics[width=0.25\textwidth]{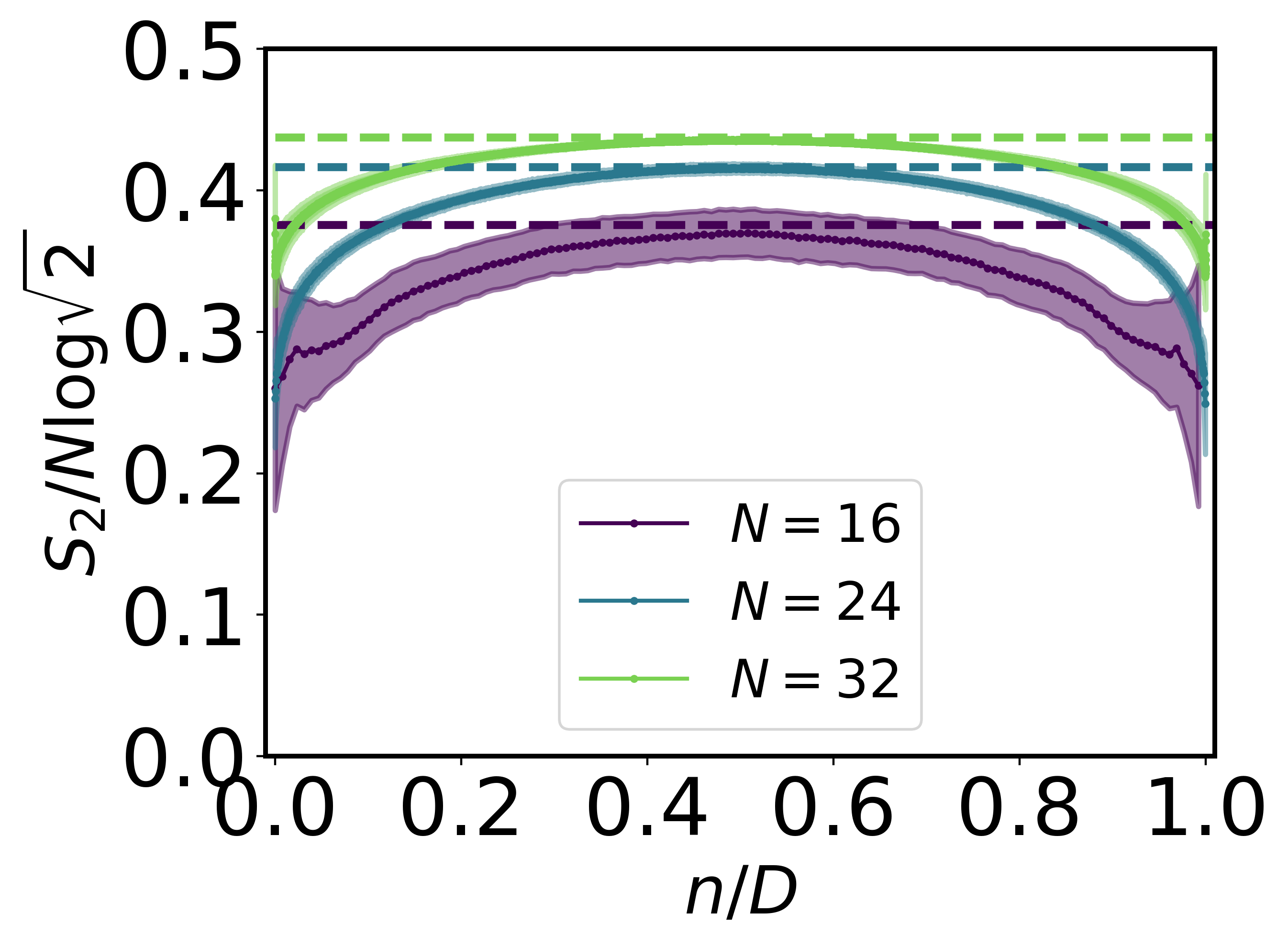}
  }%
  \caption{Entanglement and R\'enyi entropies between the $A$ and $B$ dots in each eigenstate with even fermion parity for the (a) $r=4$ and (b) $r=2$ models with $V=0.35$ and system sizes $N=16,24,32$, averaged over $1600$, $320$, and $25$ disorder realizations, respectively. 
  The solid lines represent the entropy averaged over $R$ disorder realizations and the shaded regions one standard deviation. The dashed horizontal lines indicate the Page value at the corresponding system size. 
  The horizontal axis is labelled by $n/D$, where $D$ is the Hilbert space dimension and $n$ indexes the eigenstates in order of increasing energy. \label{fig:estate-entanglement} }
\end{figure}

As a measure of the thermal character of the eigenstates of these models, 
we next compute the entanglement and R\'enyi entropies between the two dots for each eigenstate of fixed even total fermion parity. The results are plotted in Fig. \ref{fig:estate-entanglement}. 
For states near the middle of the spectrum, being effectively at infinite temperature, we expect their subsystem entanglement to be close to that of a random pure state by ETH. For the von Neumann entropy, this value is given by the Page value \cite{Page1993},
\begin{align}
    S_{A,\text{Page}} = \log D_A - \frac{D_A}{2D_B}, \label{eqn:page-value}
\end{align}
where $D_{A,B}$ are the Hilbert space dimensions of subsystem $A$ and its complement $B$. Likewise, the average R\'enyi entropy in a random pure state is given by \cite{Lubkin1978}
\begin{align}
    S_{A,\text{Page}}^{(2)} = \log\left( \frac{D_A D_B + 1}{D_A D_B} \right). \label{eqn:page-value-renyi}
\end{align}
For the $r=2$ model, $D_{A,B} = 2^{N/4}$, while $D_{A,B} = 2^{N/4-1}$ for the $r=4$ model, as each eigenstate also has definite fermion parity on each dot; hence each eigenstate only has support on half the Hilbert space of each dot. 

We see from Fig. \ref{fig:estate-entanglement} that that the entanglement entropy of states near the middle of the spectrum for both models remain close to the Page value for each system size $N$, indicating volume law entanglement. Additionally, both the entanglement and R\'enyi entropies vary continuously with the energy density, and so we do not appear to observe violations of ETH. 
We note, however, that the two models exhibit distinct behavior near the edges of the spectrum.
In the $r=2$ model, the entanglement of the lowest and highest energy eigenstates remains high. 
This is not surprising, as the quadratic interdot coupling term is more relevant (in the sense of the renormalization group) than the quartic intradot coupling, and hence will dominate the low energy physics. 
In contrast, for the $r=4$ model the entanglement drops off considerably as one approaches the edges of the spectrum. 
As $V<J$ and the interdot coupling is marginal, we expect the intradot to dominate, resulting in lower interdot entanglement at low energies.
Thus, all we can conclude here is that the eigenstate entropies in each model are consistent with ETH. In conjunction with the SFF results, we thus do not expect to see non-thermal late-time behavior in either coupled model.

\section{Thermalization in Finite-N Quenches \label{sec:finite-N-quenches} }

\subsection{ R\'enyi and Entanglement Entropy \label{sec:finite-N-quench-entanglement} }

\begin{figure}
  \subfloat[$r=4$]{
  \includegraphics[width=0.25\textwidth]{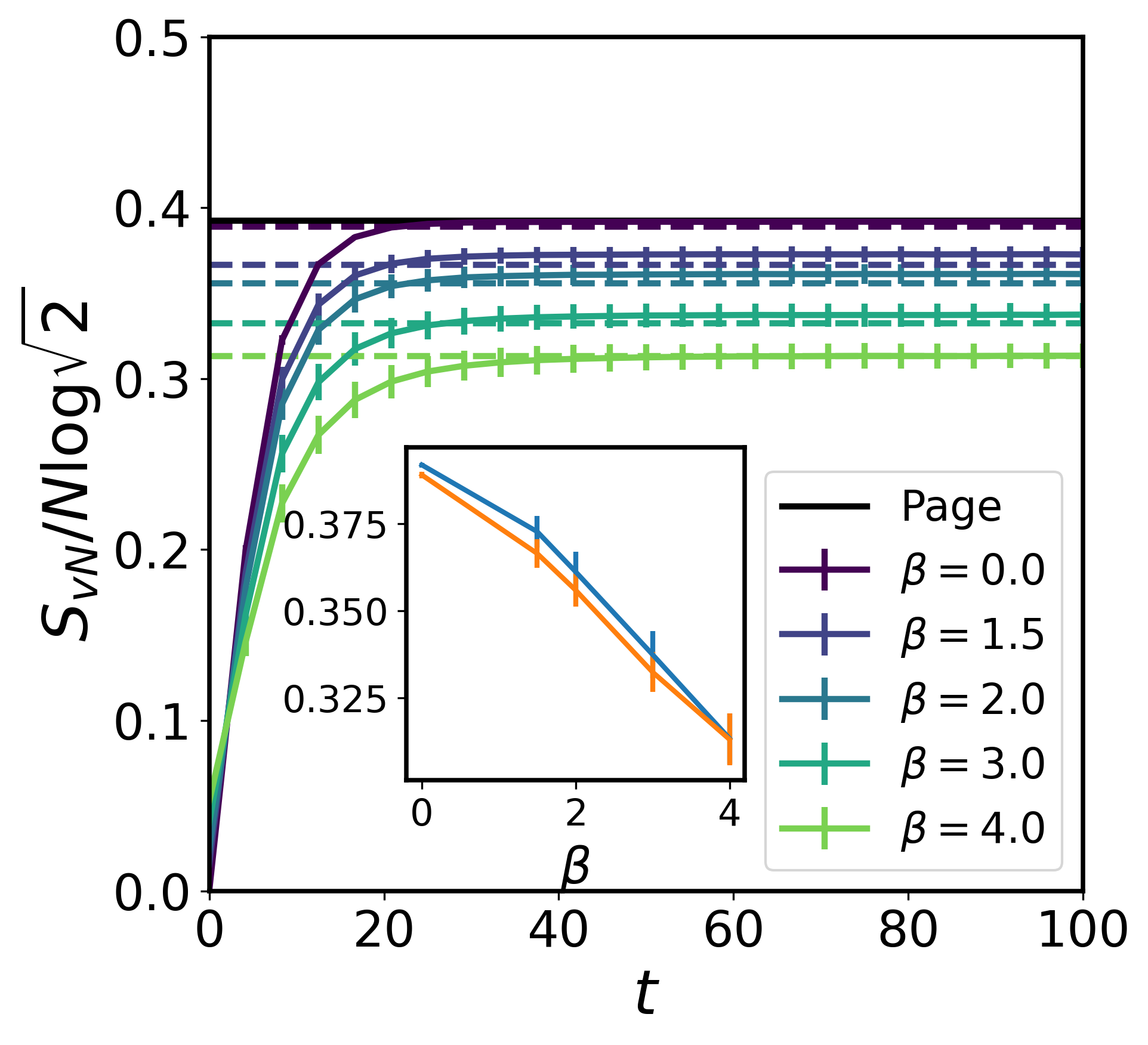}%
  \includegraphics[width=0.25\textwidth]{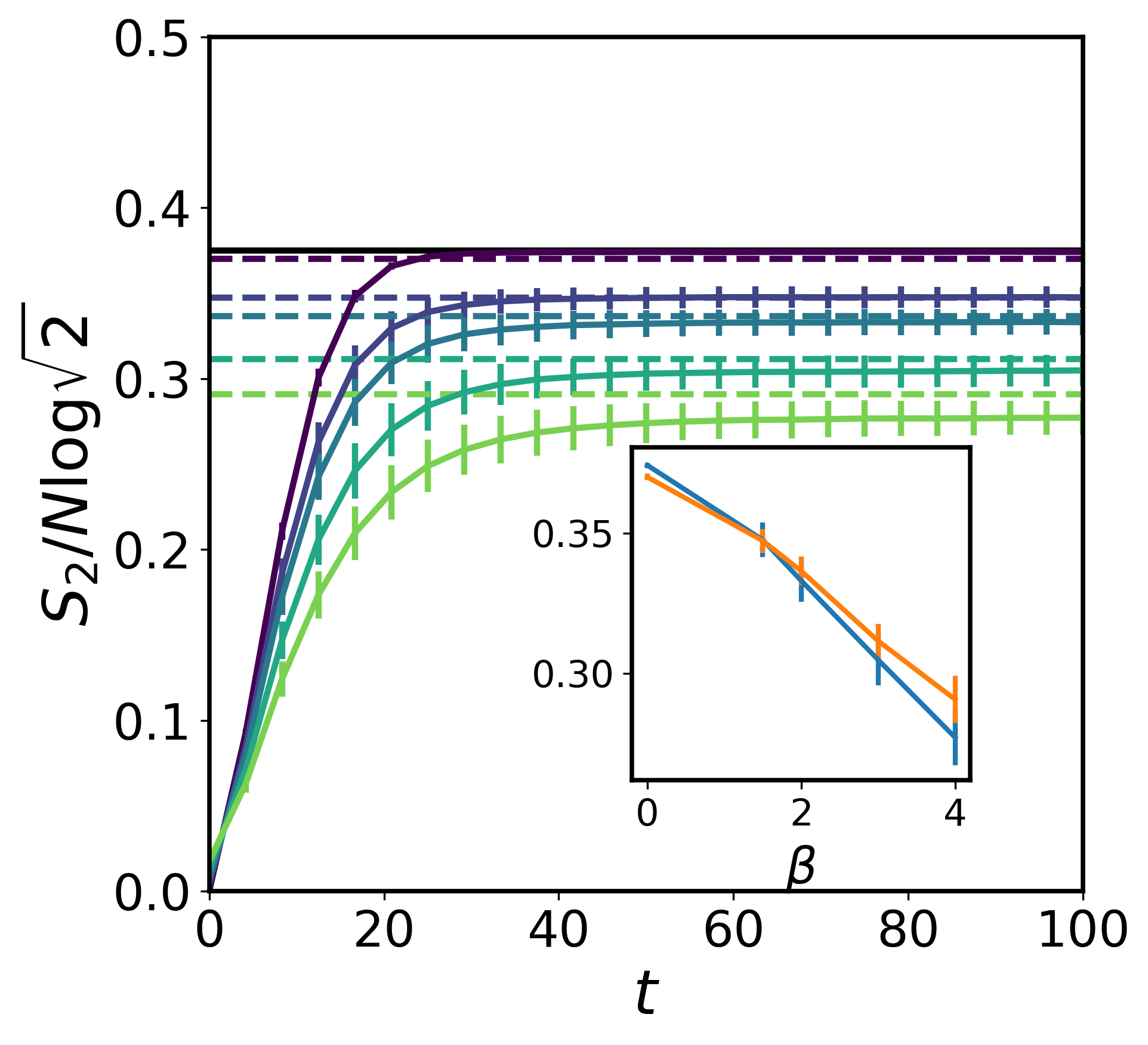}%
}%
  \subfloat[$r=2$]{%
  \includegraphics[width=0.25\textwidth]{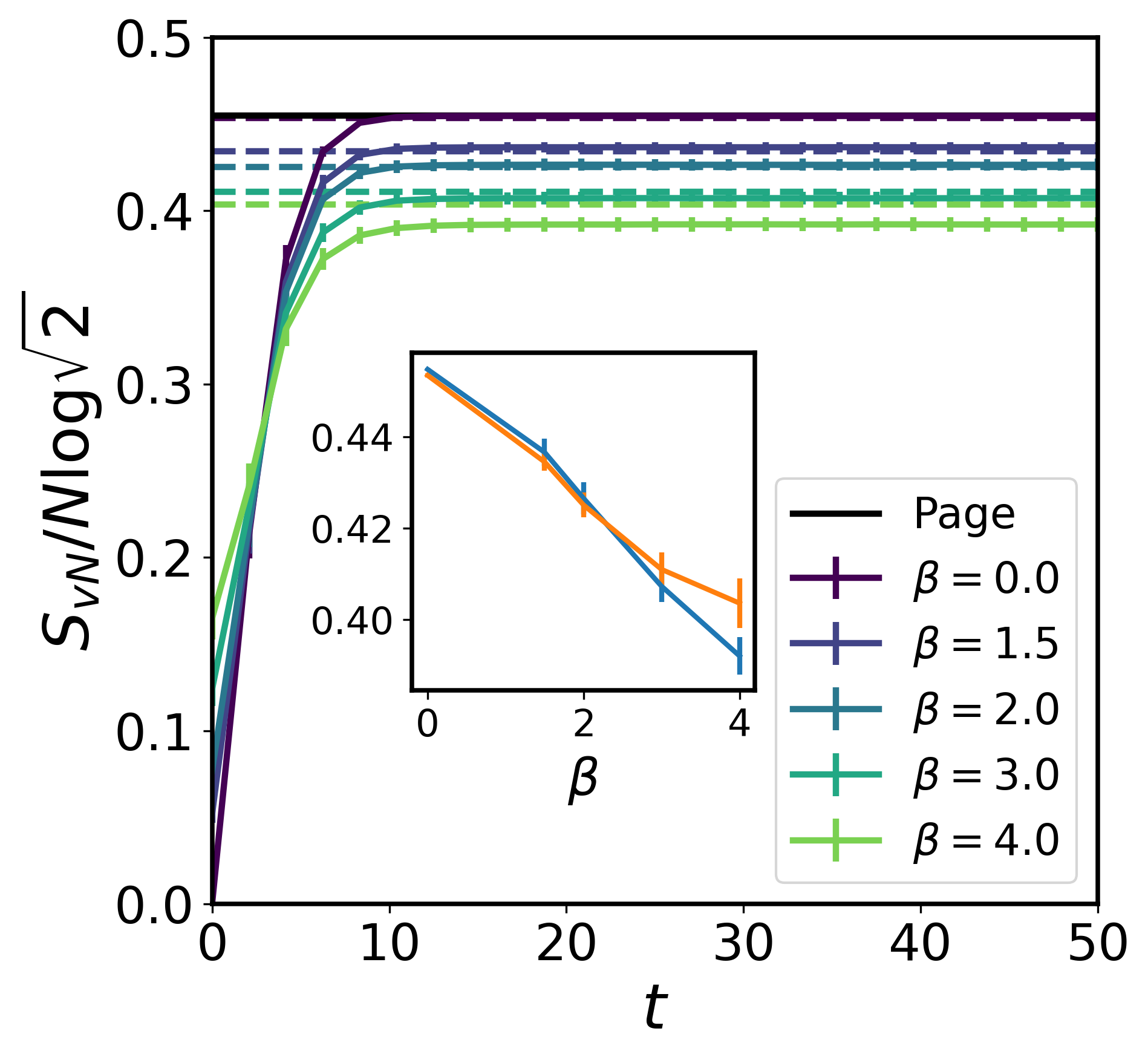}%
  \includegraphics[width=0.25\textwidth]{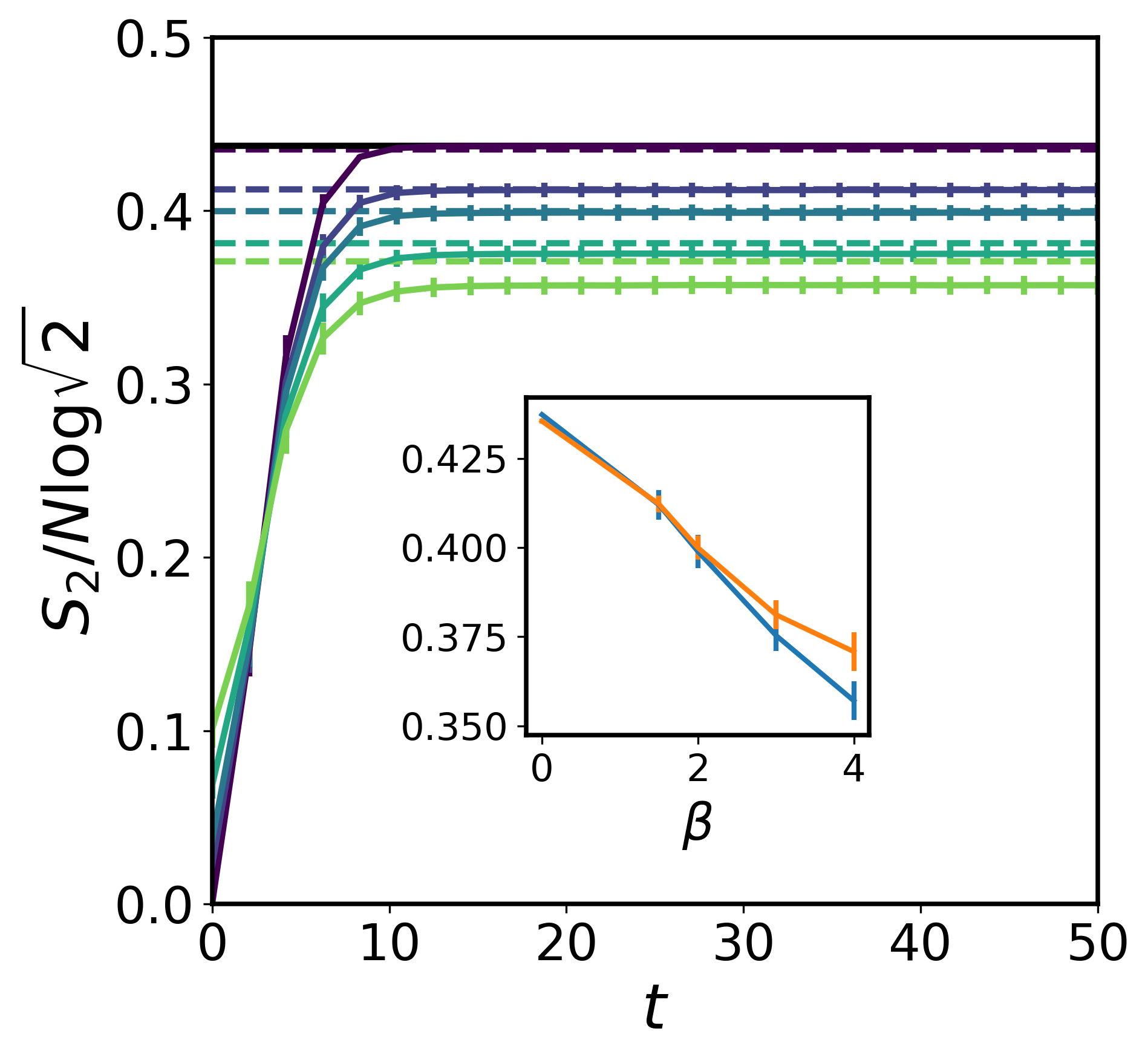}%
}
 \caption{Entanglement and R\'enyi entropies after quenches from KM states for different values of $\beta$ in the (a) $r=4$ and (b) $r=2$ coupled models with $N=32$ and $V=0.35J$, averaged over $210$ disorder realizations. The dashed lines 
 indicate the entropy of an eigenstate with energy close to the corresponding KM state, averaged over $30$ disorder realizations. The insets plot the saturation value of the entropy (blue) against the eigenstate entropy (orange) as functions of $\beta$.  \label{fig:finiteN-KM-quench} } 
\end{figure}

We now move on to reexamining the late-time entanglement of the coupled SYK models after a KM state quench. 
Again, we fix $V=0.35J$. For initial states at effectively infinite temperature, 
the expected saturation value of the entanglement and R\'enyi entropies are given by the corresponding Page values, of Eqs. \eqref{eqn:page-value}-\eqref{eqn:page-value-renyi}. For initial states with $\beta > 0$, we will test for thermalization by comparing the saturation value with the entanglement entropy of an eigenstate with (nearly) the same energy. Here we are appealing to ETH and assuming that observables computed in an eigenstate take values representative of those in the microcanonical ensemble at the corresponding eigenenergy; we believe this to be reasonable based on the eigenstate entanglement discussed above. 
Since the KM states have definite fermion parity, we choose reference eigenstates with fermion parity equal to the initial state (which ensures the $r=4$ eigenstates have fixed fermion parity on each dot).

In Fig. \ref{fig:finiteN-KM-quench} we plot the disorder averaged entanglement and second R\'enyi entropies after a quench from a KM state in both coupled SYK models for various $\beta$ and $N=32$. 
In both models, we see that the entropies saturate to the expected values for $\beta < 4$. For  $\beta=4$, the R\'enyi entropy in the $r=4$ model and both entropies in the $r=2$ model saturate slightly below the expected values. 
This is likely a finite-size effect, since, comparing with Fig. \ref{fig:estate-entanglement}, we see that these entropy values correspond to states near the edge of the spectrum, where the density of states is lower and hence the nearest reference state may be far away enough in energy to not accurately represent the microcanonical ensemble centered at the KM state energy. We expect this deviation to be reduced for larger $N$. 
Nevertheless, since the deviation is relatively small, we conclude that both coupled SYK models appear to always thermalize for the values of $\beta$ considered.

\begin{figure}
  \subfloat[$r=2$]{%
  \includegraphics[width=0.45\columnwidth]{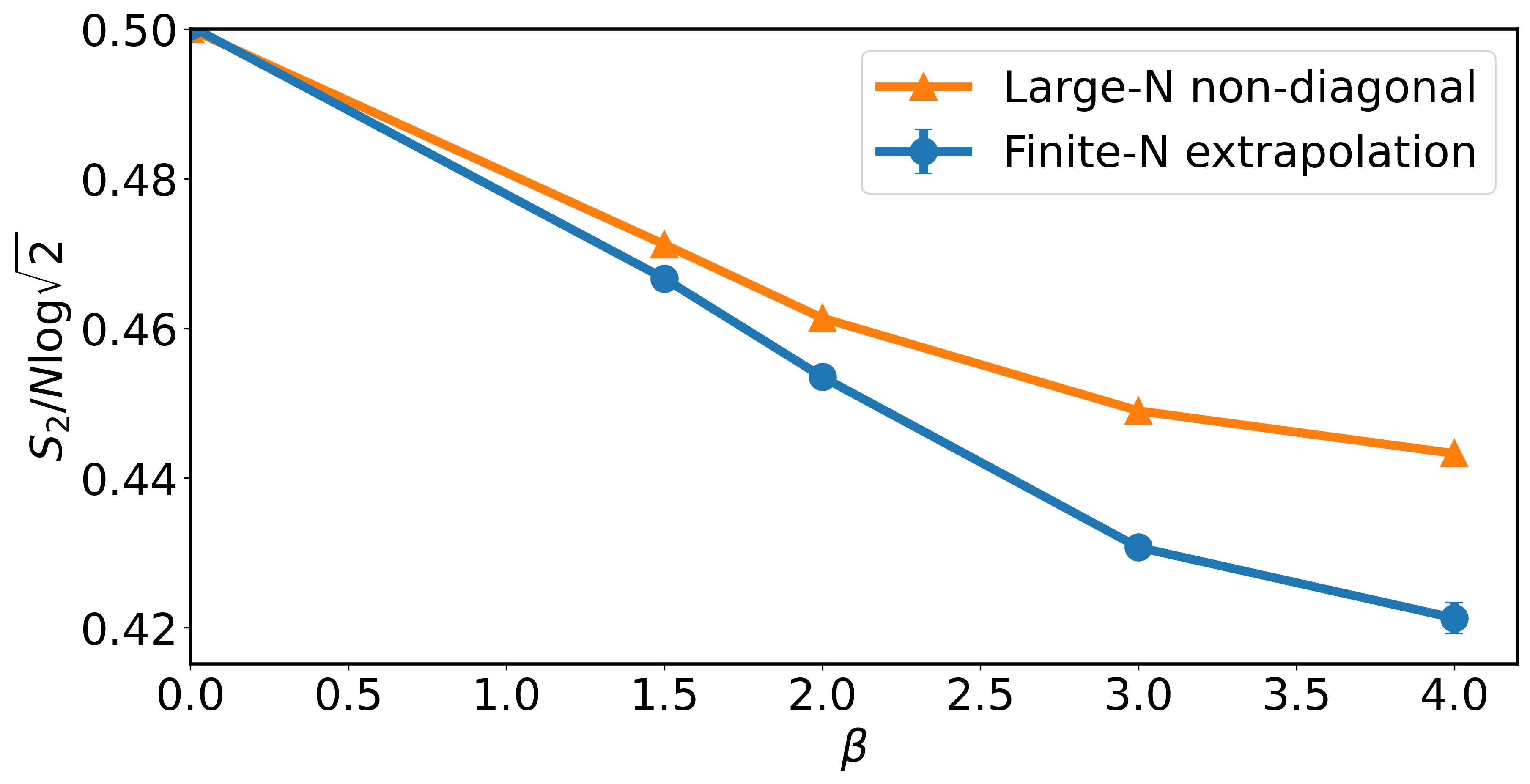}
}%
  \subfloat[$r=4$]{%
  \includegraphics[width=0.45\columnwidth]{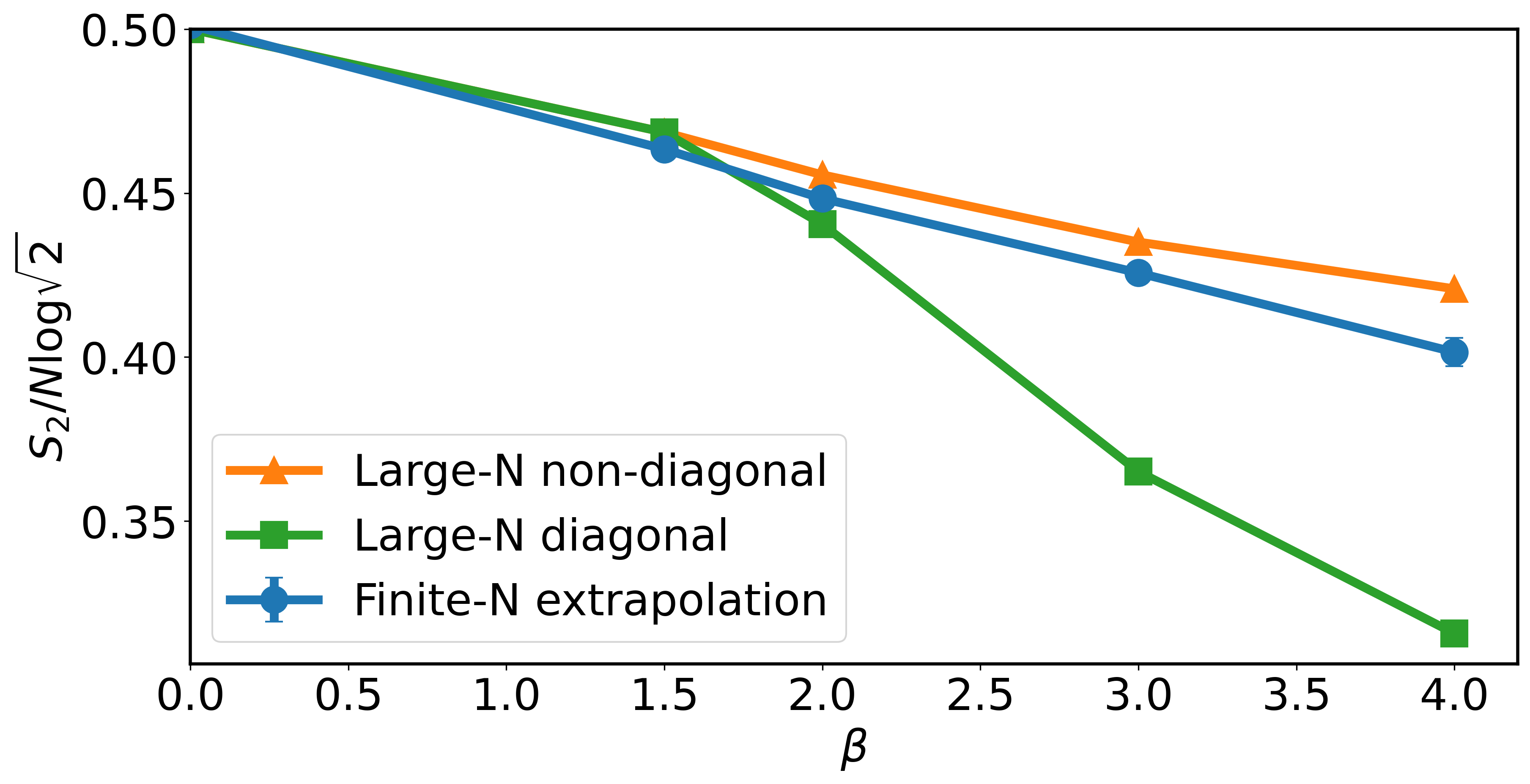}
}
  \caption{Extrapolated values of the saturation values of the second R\'enyi from finite-$N$ numerics compared against large-$N$ saddle point results as functions of $\beta$ for the (a) $r=2$ and (b) $r=4$ models with $V=0.35J$. The error bars are obtained from the covariance matrix for the linear fits and are negligble. The fits were performed using data from system sizes
  $N=24-40$ and $N=24-36$ for $r=2$ and $r=4$, respectively.
  } \label{fig:finiteN-KM-quench-extrapolation} 
\end{figure}

We next perform an extrapolation of the finite-$N$ data to $N \to \infty$ to compare with our large-$N$ results. Specifically, we linearly fit the saturation values of the R\'enyi entropies as functions of $1/N$ for each $\beta$ to extract the $N \to \infty$ result. For each $\beta$ and $N$, we average over enough disorder realizations such that the saturation value does not change appreciably with additional realizations. In Fig. \ref{fig:finiteN-KM-quench-extrapolation} 
we plot the results of the linear extrapolations compared against the large-$N$ saturation values after quenches from KM states. 
In the case of the $r=2$ model, we see that the extrapolated data agrees relatively well with the large-$N$ results. The deviation between the two grows with increasing $\beta$, which we again attribute to finite size effects.
Turning to the $r=4$ model,
in Fig. \ref{fig:finiteN-KM-quench-extrapolation}(b), we plot both the large-$N$ replica non-diagonal saddle-point as well as the diagonal saddle-point when the system does not thermalize.  
Although our large-$N$ numerics predicted subthermal behavior for $\beta \geq 2$ we see that the finite-$N$ data appears to extrapolate most closely to the replica non-diagonal saddle-point, which would suggest the system \emph{thermalizes} in the large-$N$ limit.

Based on the R\'enyi and entanglement entropies, our exact diagonalization results thus all imply thermalization at finite  $N$ and appear to predict thermalization in the limit $N \to \infty$. 
One must of course take the extrapolation of our finite-$N$ data with a grain of salt, given the small system sizes to which we have access. But taken at face value, our data seems consistent with the interpretation of Ref. \cite{Gu2017b}, namely that heavy modes, which contribute to the zero-temperature thermal entropy of the $r=4$ model, take an infinite time to thermalize in the large-$N$ limit. At fixed, finite values of $N$, there is of course no zero-temperature entropy and hence one would not expect to see any slowly thermalizing degrees of freedom. One might then expect that the large-$N$ subthermal behavior might give way to thermalization when 
subleading corrections are taken into account. We will speculate on this briefly in Section \ref{sec:discussion}.  

\subsection{Reduced Density Matrix Spectral Form Factor \label{sec:rdm-sff} }

The results of the preceding section tell us that the subthermal behavior observed at large $N$ for the $r=4$ coupled model does not appear to persist down to finite $N$. However, entanglement entropy, as a single number, provides a very coarse characterization of the reduced density matrix (RDM). Indeed, by examining the full \emph{entanglement spectrum}, one might hope to find some difference between how the coupled SYK models thermalize. To that end, we proceed to compute the spectral form factor (SFF) of the RDM after equilibrium is reached following a quench from a KM state. Following Ref. \cite{Chen2018}, we consider the singular values $\lambda_i$ of the RDM $\rho_A$ and define the RDM SFF as
\begin{align}
    g(\tau) = \langle \sum_{i,j} e^{i\tau(\lambda_i - \lambda_j)} \rangle,
\end{align}
where $\tau$ is a fictitious time. We can also define a connected RDM SFF as
\begin{align}
    g_c(\tau) = g(\tau) - |\langle \sum_i e^{i\tau\lambda_i} \rangle|^2,
\end{align}
which subtracts out non-universal contributions at small values of $\tau$. Akin to the SFF of the Hamiltonian, the RDM SFF provides a probe of repulsion between the singular values of the RDM.

Now, as argued in Ref. \cite{Chen2018}, the RDM SFF of a state obtained by evolving an initially short-range entangled state with a chaotic Hamiltonian will display universal behavior governed by the appropriate Wishart random matrix ensemble (to be reviewed below) including the standard slope, ramp, and plateau features. 
Indeed, ETH implies that the reduced density matrix should take the form $\rho_A \sim e^{-H_A}$, where $H_A$ is the physical Hamiltonian projected to subsystem $A$ \cite{Garrison2018}. So, if $H$ describes a chaotic system and hence has level statistics described by random matrix theory, one expects the same to be true for $\rho_A$. Thus, computing the RDM SFF amounts to a finer test of ETH. 

Now, when quenched from a pure state with infinite effective temperature, we expect a chaotic many-body system to equilibrate to a random state. Bipartionining the Hilbert space into subregions $\mathcal{H}_A$ and $\mathcal{H}_B$, we can write such a random state in a Schmidt-decomposed form:
\begin{align}
    \ket{\Psi} = \sum_{i=1}^{D_A}\sum_{J=1}^{D_B} X_{iJ} \ket{\Psi_A^i}\ket{\Psi_B^J},
\end{align}
where $D_{A,B}$ are the dimensions of Hilbert spaces $\mathcal{H}_{A,B}$, $\ket{\Psi_{A,B}^{i,J}}$ are complete bases of states for each subregion, and $X_{iJ}$ are random complex Gaussian variables, subject to the normalization constraint $\Tr[XX^\dagger] = 1$. The RDM for region $A$ is then simply given by $\rho_A = XX^\dagger$. We can model such a distribution of random RDMs using the \emph{Wishart-Laguerre} random matrix ensemble \cite{Liu2018sff,edelman_rao_2005,forresterbook}. Indeed, if we consider an unconstrained $D_A \times D_B$ matrix $Y$, with independent complex entries sampled from the Gaussian distribution
\begin{align}
    P(\{Y_{iJ}\}) = \mathcal{N}^{-1} \exp\left( - D_B \Tr(YY^\dagger) \right),
\end{align}
then the matrix $W=YY^\dagger$ belongs to the unitary Wishart random matrix ensemble. The RDM for the random Page state can then be constructed as 
\begin{align}
    \rho_A = \frac{YY^\dagger}{\Tr[YY^\dagger]}.
\end{align}
In the limit $D_{A,B} \to \infty$ with the ratio $D_A / D_B$ held fixed, the denominator has mean $D_A$ with vanishing fluctuations, and so we can write the RDM in terms of an unconstrained matrix drawn from the Wishart ensemble: $\rho_A  = YY^\dagger / D_A$ \cite{Chen2018}. The SFF computed from the Wishart ensemble thus serves as a benchmark for a maximally random RDM.

We now proceed to an analysis of the late-time connected RDM SFFs\footnote{We focus on the connected RDM SFF, as the ramp is suppressed in the full RDM SFF for an equal bipartition of the Hilbert space \cite{Chen2018}.} for both coupled SYK models 
for different values of $\beta$ and system sizes $N$. For comparison, we also compute the RDM SFF for the single SYK model (the subsystems $A$ and $B$ correspond to an arbitrary, equal bipartitioning of the fermions). Now, we recall that since the KM state is a state of definite fermion parity on both subsystems $A$ and $B$ and only the $r=4$ coupled SYK Hamiltonian conserves fermion parity separately on both subsystems, the effective Hilbert space dimension involved in the time evolution in the KM quench is different between the three models. For ease of comparison, we will thus consider quenches from the indefinite fermion parity state $\ket{\widetilde{KM}_\beta}$ of Eq. \eqref{eqn:KM-indefinite-parity}, which mixes all parity sectors and results in time evolution involving the full Hilbert space in all three models, in this section. 

\subsubsection{\texorpdfstring{$\beta = 0$}{beta=0}}

\begin{figure*}
  \subfloat[$r=4$]{%
  \includegraphics[width=0.32\textwidth]{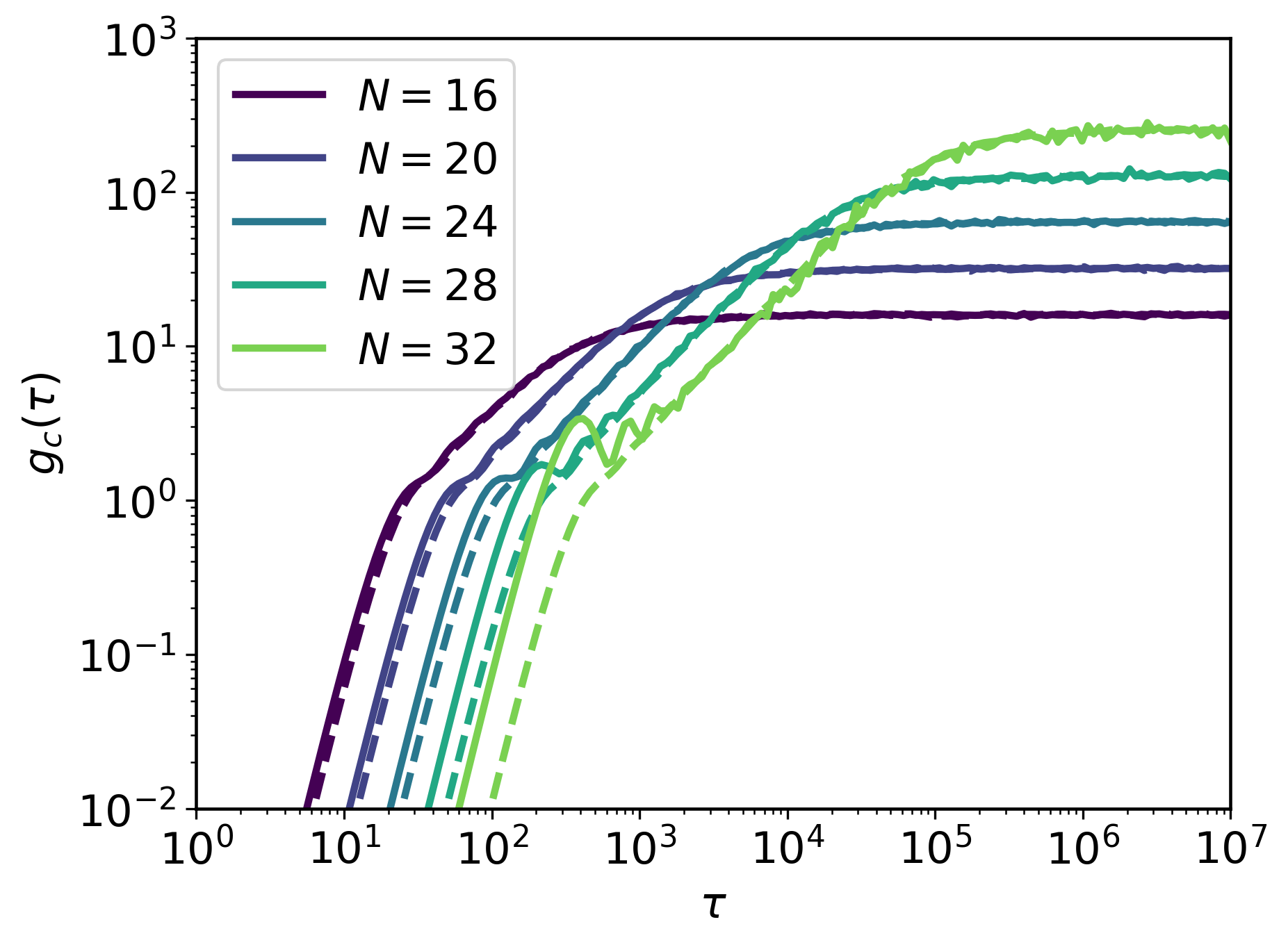}%
}
  \subfloat[$r=2$]{%
  \includegraphics[width=0.32\textwidth]{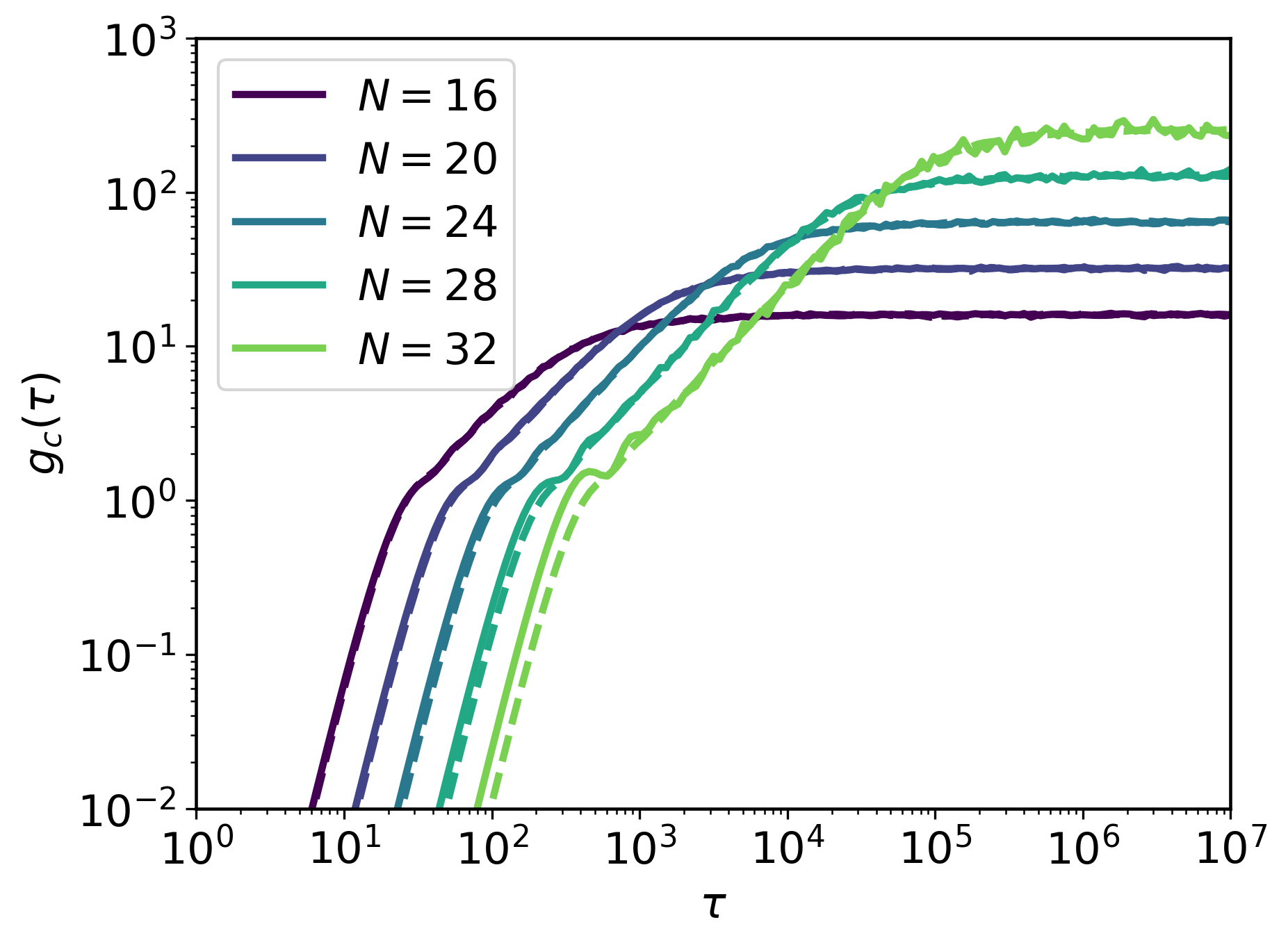}%
}
  \subfloat[single]{%
  \includegraphics[width=0.32\textwidth]{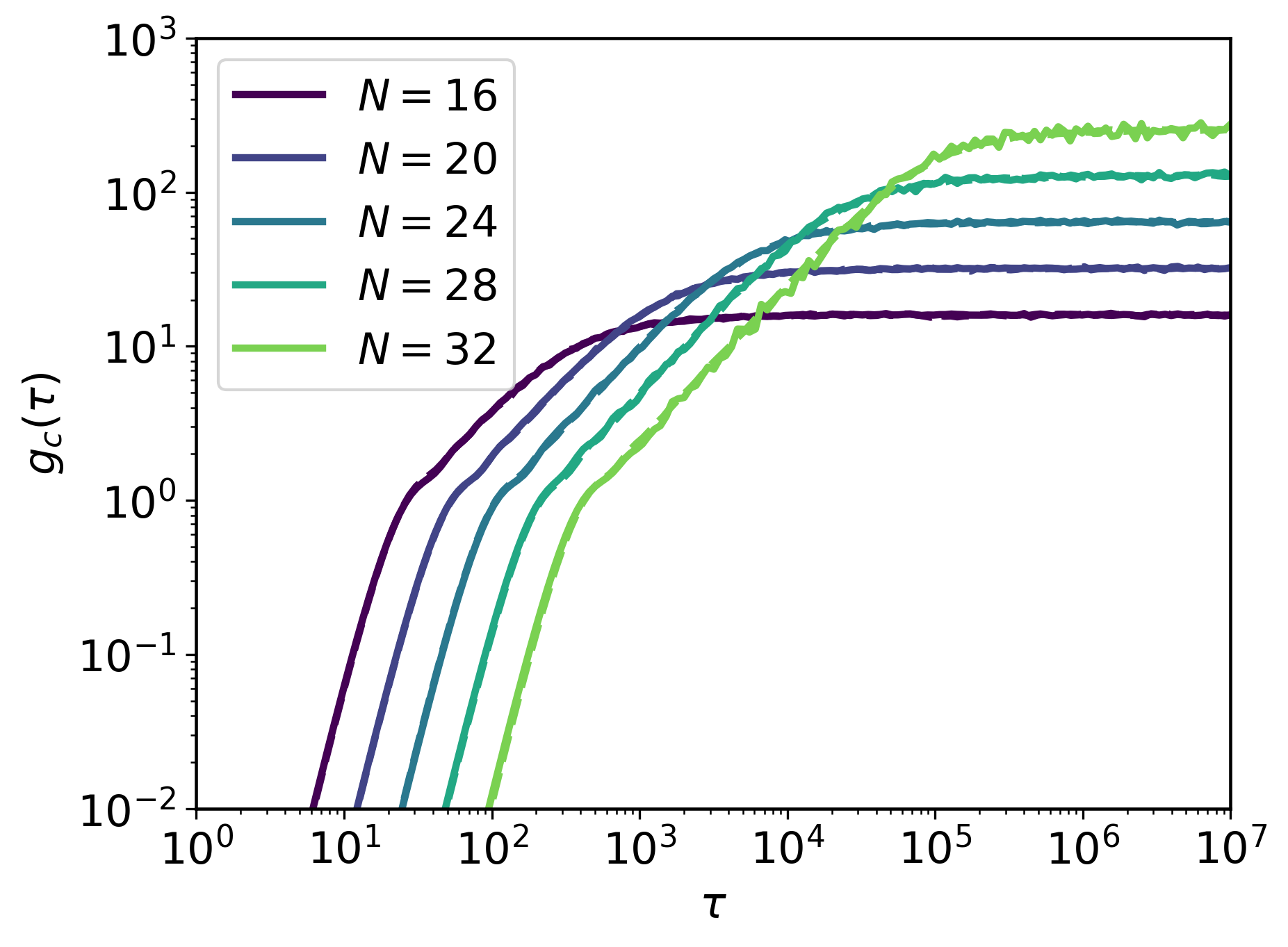}%
}
  \caption{Late-time connected RDM spectral form factors after a quench from the $\ket{\widetilde{KM}_{\beta = 0}}$ state of Eq. \eqref{eqn:KM-indefinite-parity} in the (a) $r=4$, (b) $r=2$, and (c) single SYK models for $N=16,20,24,28,32$, with the ensemble averages taken over $16000,16000,3200,960,210$ disorder realizations, respectively. 
  The dashed lines indicate the Wishart ensemble SFF for the corresponding system size. }
  \label{fig:rdm-sff-beta-0}
\end{figure*}

The results for a $\beta=0$ quench are depicted in Fig. \ref{fig:rdm-sff-beta-0}, with the random matrix theory (RMT) predictions presented as dashed lines. 
We see that the late-time RDM SFFs of all three models display clear slope-ramp-plateau structures. 
In particular, the single SYK RDM SFF exhibits perfect agreement with RMT. 
In contrast, while the RDM SFFs of the coupled SYK models agree well with RMT, they exhibit a slight overshoot where the slope joins the ramp, which grows with system size and is more pronounced in the $r=4$ model.

Let us set these observations in the context of other non-integrable models. In Ref. \cite{Chen2018}, it was found that the late-time RDM SFF of an energy non-conserving Floquet model agreed perfectly with the RMT prediction, while an energy conserving Ising model exhibited an overshoot of the ramp like that we see in the coupled model (though significantly more pronounced). The authors argued that the disagreement with RMT in the Ising model was a consequence of energy conservation acting as a constraint on the evolution of the wave function, inhibiting the development of chaos. However, the late-time RDM SFF of the single SYK model is exactly consistent with RMT, suggesting that energy conservation is not, on its own, sufficient to inhibit thermalization. Though this was not the main object of our study, we emphasize that this is an interesting result in its own right.
Indeed, the fact that the late-time RDM SFF exactly resembles that of a completely random state provides a novel characterization of the sense in which the SYK model is maximally scrambling, in spite of its energy conservation. 

The mild overshoot of the ramp in the coupled models can thus likely not be attributed only to energy conservation alone. The fact that the overshoot develops for both coupled models indicates that its origin does not lie with the fermion parity conservation in the $A$ and $B$ dots of the $r=4$ model.
The primary distinction between the coupled and single SYK models is that the former have some notion of locality -- the models describe two clusters of Majorana fermions which can be interpreted as being spatially separated -- and hence the Hamiltonian will be more sparse than that of the latter model. It thus seems reasonable to conclude that the non-local nature of the couplings in the single SYK model are what lead to the perfect agreement of the RDM SFF with the RMT prediction. In particular, the fact that similar deviations from RMT are present in both coupled models suggest that they are not linked with a potential source of subthermal behavior in the $r=4$ model at large-$N$. In fact, the small magnitude of deviations suggests that both are in fact excellent scramblers. 

\subsubsection{\texorpdfstring{$\beta \neq 0$}{beta neq 0}}

\begin{figure*}
  \subfloat[$\beta=1.5$]{%
  \includegraphics[width=0.32\textwidth]{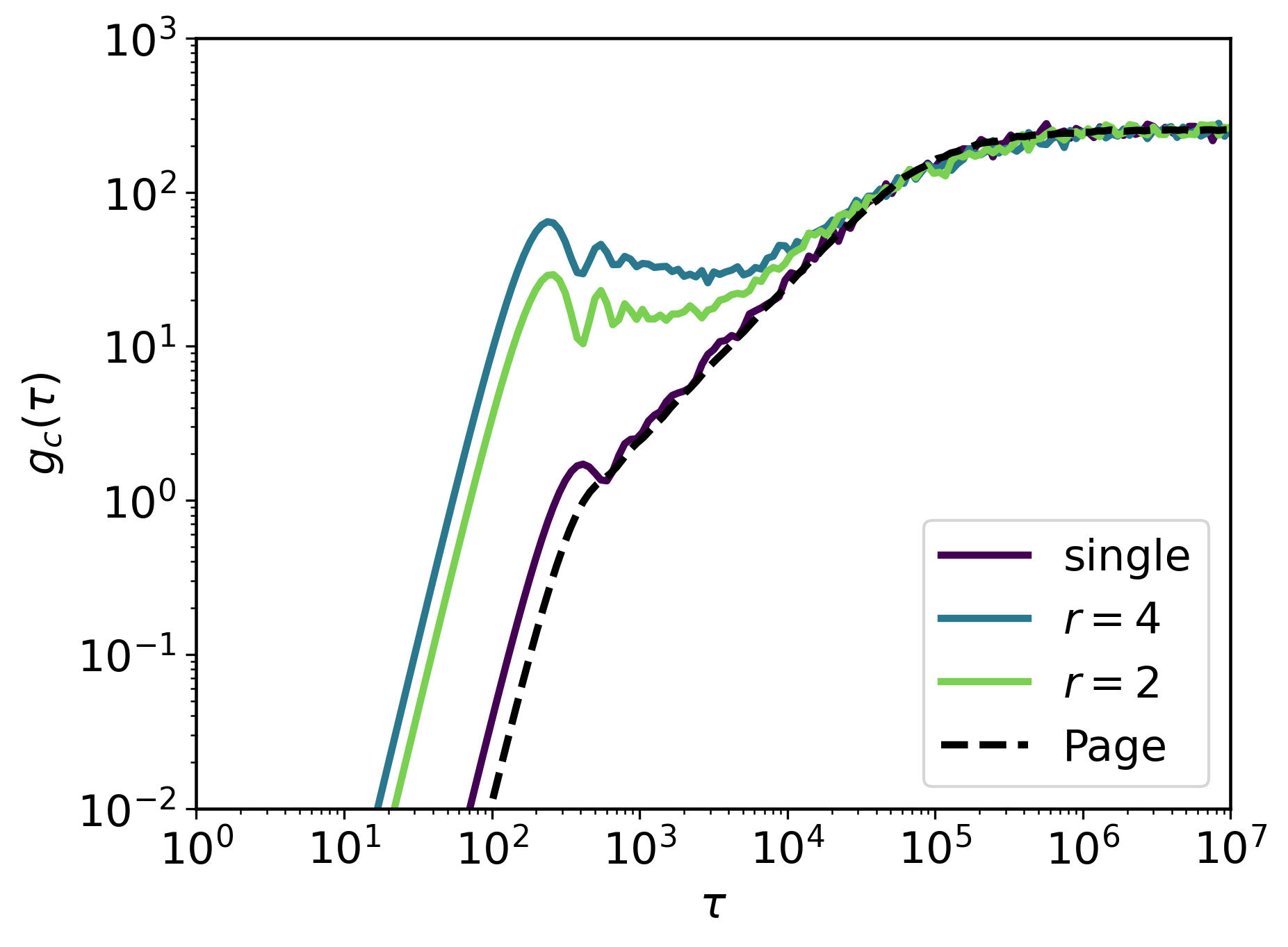}
}
\subfloat[$\beta=3$]{%
  \includegraphics[width=0.32\textwidth]{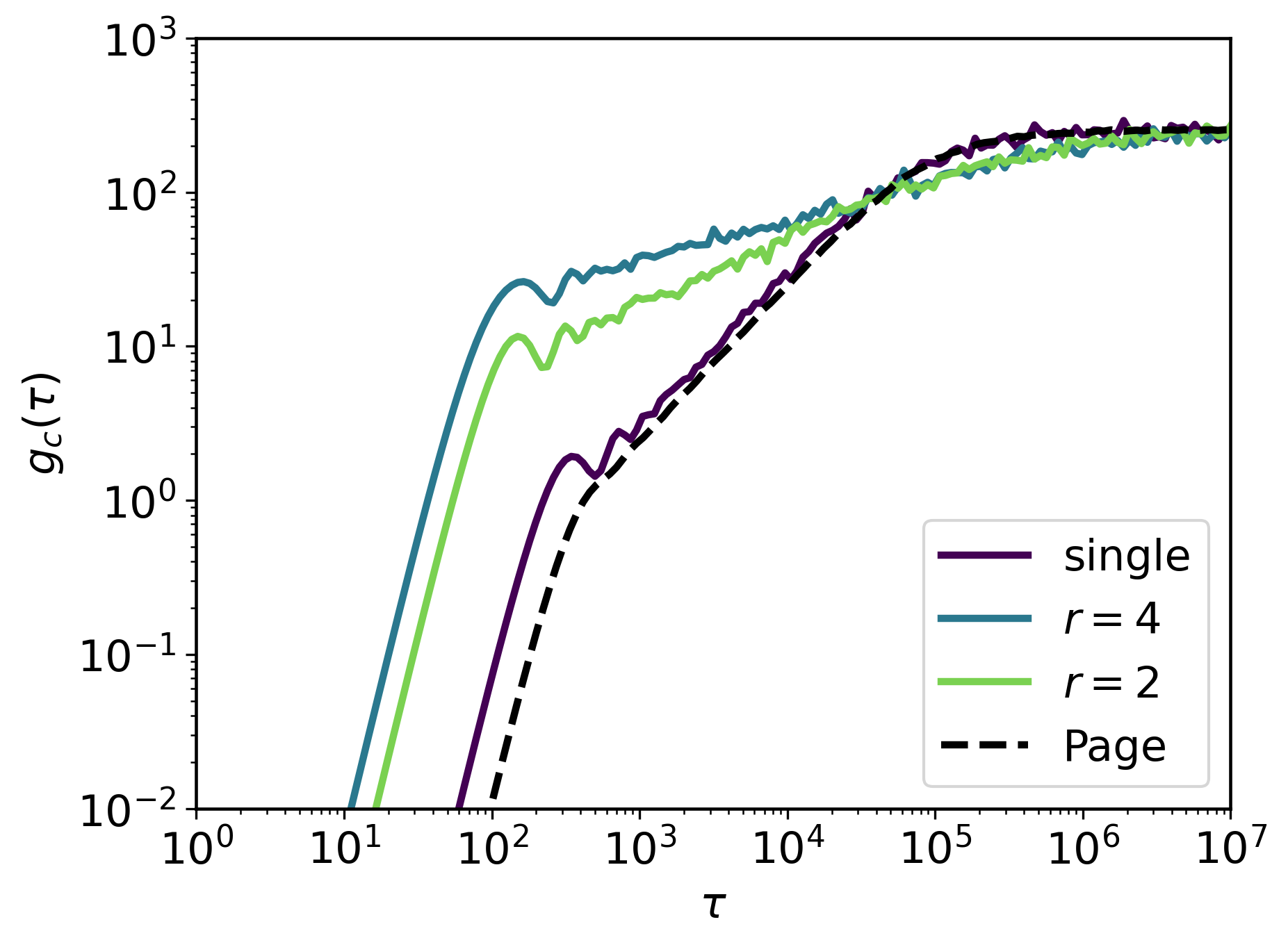}
}
\subfloat[$\beta=4$]{%
  \includegraphics[width=0.32\textwidth]{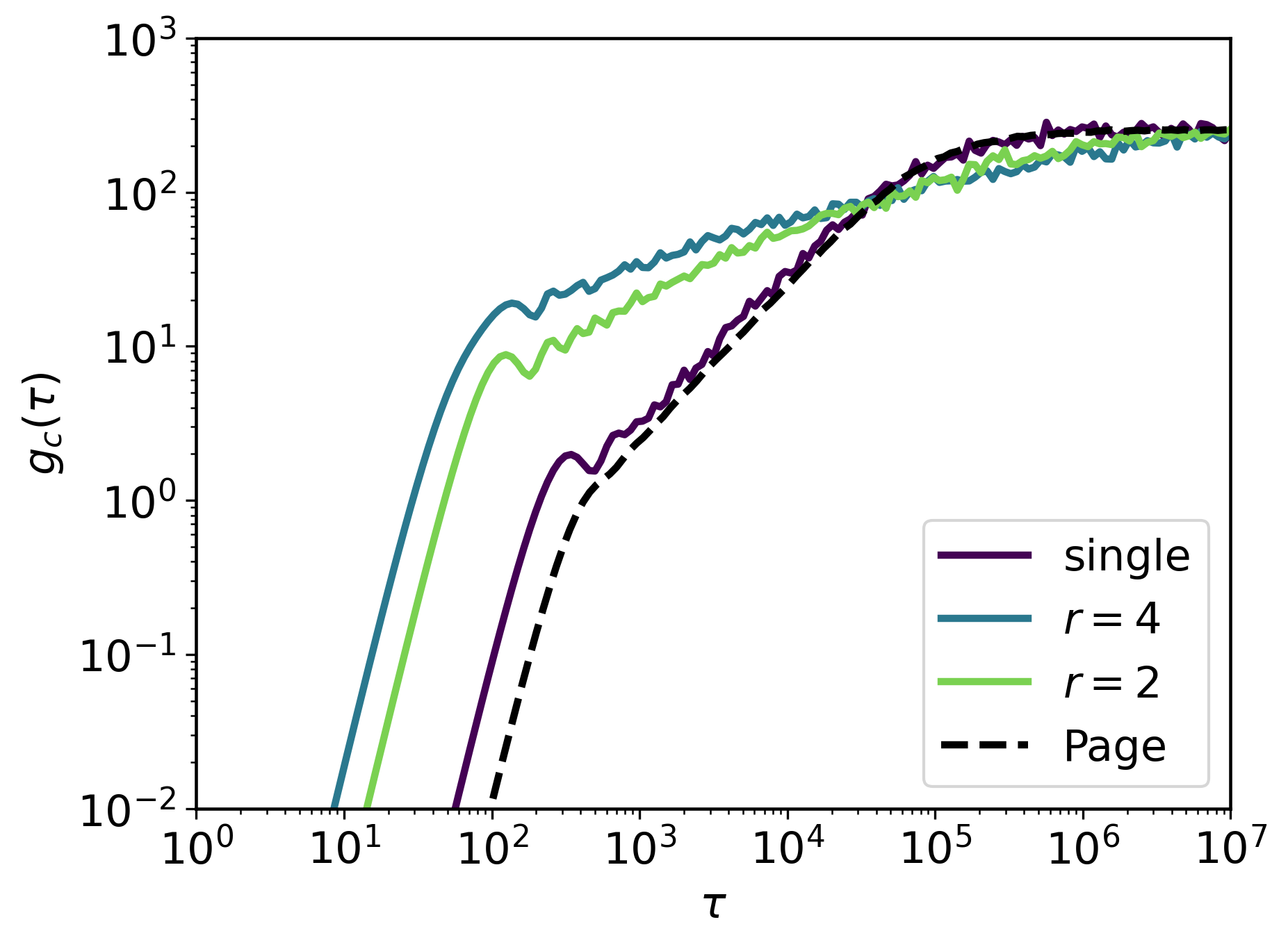}
}
  \caption{Late-time connected RDM spectral form factors after a quench from the $\ket{\widetilde{KM}_{\beta = 0}}$ state in the $r=4$, $r=2$, and single SYK models for different values of $\beta$, averaged over $210$ disorder realizations with a system size of $N=32$, compared against the RMT prediction.}  
  \label{fig:rdm-sff-beta}
\end{figure*}

Finally, we inspect the RDM SFFs after quenches from $\ket{\widetilde{KM}_\beta}$ states with non-zero $\beta$, the data for which is shown in Fig. \ref{fig:rdm-sff-beta}. We recall that, for $\beta \geq 2$, the $r=4$ model reaches a subthermal state in large-$N$. 
Nevertheless, the connected RDM SFFs of both the $r=2$ and $r=4$ models exhibit similar behavior. 
In particular, already at $\beta = 1.5$, the RDM SFFs exhibit significant departures from a linear ramp for smaller $\tau$, while there is no vestige of linear growth for any $\tau$ at larger values of $\beta$.
In contrast, the RDM SFF of the single SYK model still exhibits a clear linear ramp, indicating the retention of spectral rigidity in the RDM spectrum. Evidently the single SYK model remains an excellent scrambler over a large temperature range.

Now, we emphasize that the definition of $\ket{\widetilde{KM}_\beta}$ for finite $\beta$ depends on the choice of Hamiltonian, so we cannot directly compare the RDM SFFs of the three models, as they have been quenched from different states. Additionally, since we are now quenching from states which do not correspond to infinite temperature, 
we should not necessarily expect \textit{a priori} that the system should equilibrate to a completely random state. However, we take the fact that the RDM SFFs of the $r=2$ and $r=4$ models are at least qualitatively similar as further evidence of the claim that the origin of the subthermal behavior of the $r=4$ model observed in large-$N$ is an artifact of that limit and does not manifest at finite $N$.

\section{Discussion and Conclusion \label{sec:discussion} }

We have presented a thorough characterization of the post-quench entanglement of randomly coupled SYK models, motivated by the analytic calculations of Ref. \cite{Gu2017b}, which found subthermal behavior in the large-$N$ limit after a quench from a TFD state in the large effective temperature limit. In our large-$N$ analysis, we expanded on previous analytic and numerical results and illustrated that the $r=4$ coupled SYK model exhibits state-dependent thermalization for sufficiently strong interdot coupling, while the $r=2$ model appears to always thermalize (at least for the parameter ranges considered). We then proceeded to characterize these coupled systems at finite-$N$ using exact diagonalization. We did not find concrete signatures of the large-$N$ subthermal behavior in these finite-size systems. Instead, they appeared to be chaotic, as characterized by the spectral form factor, and both the R\'enyi and von Neumann entropies saturated to the expected thermal values after quenches.

As a finer-grained characterization of thermalization and chaos, we computed the SFF of the late-time RDM after these quenches, comparing the results with the standard SYK model. We found that for $\beta = 0$ quenches, corresponding to infinite effective temperature, the single SYK model agreed perfectly with RMT predictions while the two coupled models exhibited slight deviations. For $\beta \neq 0$ quenches, the single SYK still agreed well with RMT, while the coupled models exhibited greater deviations. Given the similar behavior of the coupled SYK models, we could not connect these deviations to an origin for the large-$N$ subthermal behavior. However, a side result of this analysis was another characterization of the sense in which the single SYK model is a perfect scrambler. We thus conclude that the subthermal behavior of the $r=4$ model in the large-$N$ limit is an artifact of said limit.

Though our analysis is consistent with the conjecture in Ref. \cite{Gu2017b} that the subthermal value of the late-time R\'enyi entropy is a consequence of  heavy degrees of freedom which are frozen/localized in the large-$N$ limit, it remains to be seen whether this picture can be made more precise. In particular, it would be interesting to see whether taking into account subleading corrections of the large-$N$ calculation would show how a late-time thermal state emerges. It is tempting to speculate that tunneling between the replica diagonal and non-diagonal saddle-points, described by non-perturbative instanton processes, may lead to the expected finite-$N$ thermalization. Instantons in large-$N$ (Brownian) SYK systems have been discussed, but in the separate physical context of measurement induced purification transitions \cite{Bentsen2021}. It would also be interesting to see whether the state-dependent thermalization of the $r=4$ model may be addressed within a holographic computation \cite{Ryu2006a,Ryu2006b,Hubeny2007,Dong2016}.

We have focused our attention on a system of two coupled SYK dots, 
however, this two-site model was studied in Ref. \cite{Gu2017b} as a simple limiting case of a full SYK chain, which exhibits the same subthermal properties at large-$N$. The numerical methods employed here can be used to investigate entanglement propagation in the full SYK chain in the large-$N$ limit (as has already been done in the context of measurement-induced transitions in SYK chains \cite{Liu2020}). In particular, it would be of interest to better understand the potentially sub-linear scaling of R\'enyi entropy in the time regime identified in Ref. \cite{Gu2017b} following the initial linear growth. Understanding how the membrane picture \cite{Nahum2017,Nahum2018,Jonay2018,vonKeyserlingk2018,Zhou2020}, which has emerged as a means of understanding entanglement growth in chaotic systems via a space-time minimal surface determined by the subregion geometry, should be modified to account for potentially sublinear entanglement growth and subthermal saturation presents an open intriguing problem. 
We anticipate investigating the entanglement dynamics of such SYK chains in future work.


\section*{Acknowledgements}
We thank D. Abanin for a helpful discussion as well as X. Chen and Y. Gu for helpful comments. The exact diagonalization calculations were performed using the \textsc{QuSpin} package \cite{Quspin1,Quspin2}. RS acknowledges the support of the Natural Sciences and Engineering Research Council of Canada (NSERC) [funding reference number 6799-516762-2018]. This work was also supported in part by the US National Science Foundation under Grant No. DMR-1725401 at the University of Illinois (EF, LN, RS), by a fellowship at the Institute for Condensed Matter Theory of the University of Illinois (LN), and by the Gordon and Betty Moore Foundation's EPiQS Initiative through the grant GBMF 8691 (XQS). This work made use of the Illinois Campus Cluster, a computing resource that is operated by the Illinois Campus Cluster Program (ICCP) in conjunction with the National Center for Supercomputing Applications (NCSA) and which is supported by funds from the University of Illinois at Urbana-Champaign.

\appendix

\section{Thermofield Double State Quenches \label{app:thermofield-double} }

In this Appendix we provide, for comparison, data for quenches from thermofield double (TFD) states, as have been considered in Refs. \cite{Penington2019,Chen2020,Jian2021}.
To define the TFD state, we start by doubling the Hilbert space such that we have two copies of the coupled SYK model, labeled by the index $b = 1,2$. We denote the fermions in this doubled system by $\chi_i^{ab}$. This pair of coupled SYK systems are decoupled between the two Hilbert spaces and are described by the Hamiltonian
	$H_D = H_1 + H_2 \equiv H_{q,r} \otimes \mathbf{1} + \mathbf{1} \otimes H_{q,r}^T$,  
where 
	$H_{q,r}^T \equiv (-1)^{q/2} H_q^A + (-1)^{q/2} H_q^B + (-1)^{r/2} H_r$.
This doubled system is illustrated in Fig. \ref{fig:tfd-setup}(a).
Next, we define the infinite temperature TFD state $\ket{I}_{12}$ via the condition
\begin{align}
	(\chi_{i}^{a1} + i \chi_{i}^{a2})\ket{I}_{12} &= 0 \quad \forall \, i, \quad a=A,B. \label{eqn:tfd-inf-temp-cnd}
\end{align}
In this state, the $\chi^{a1}$ fermions are maximally entangled with the $\chi^{a2}$ fermions, while there is no entanglement between the $\chi^{Ab}$ and $\chi^{Bb}$ fermions. 
The TFD state at inverse temperature $\beta$ is given by $\ket{TFD_{\beta}} = Z(\beta)^{-1/2} e^{-\beta H_D/4} \ket{I}_{12}$, 
where $Z(\beta) = \Tr [ e^{-\beta H_{q,r}} ]$. 
While $\ket{TFD_{\beta}}$ is an eigenstate of $H_1-H_2$, it is \emph{not} an eigenstate of the Hamiltonian $H_D$. 
Indeed, using the fact that the TFD is annihilated by $H_1-H_2$, we can write the time-evolved TFD state as
\begin{align}
	\ket{TFD_{\beta}(t)} = 
	\frac{e^{-(i2t+\beta/2)H_1}}{\sqrt{Z(\beta)}} \ket{I}_{12}.
\end{align}
Note that, on tracing out the $b=2$ fermions in the TFD state, one obtains a thermal reduced density matrix for the $b=1$ fermions: $\rho_1 \propto e^{-\beta H_{q,r}}$, and vice versa. While each side of the TFD state is thermal by construction, a subsystem containing, say, the $\chi^{A1}_i$ and $\chi^{A2}_i$ fermions need not be thermal. If the the system thermalizes after a quench from the TFD state, 
the R\'enyi entropies should saturate to values corresponding to those computed in the thermal state $\rho_{\beta} \propto e^{-\beta (H_1 + H_2)}$. 

\subsection{Large-$N$ Quenches}

\begin{figure}
  \includegraphics[width=0.7\textwidth]{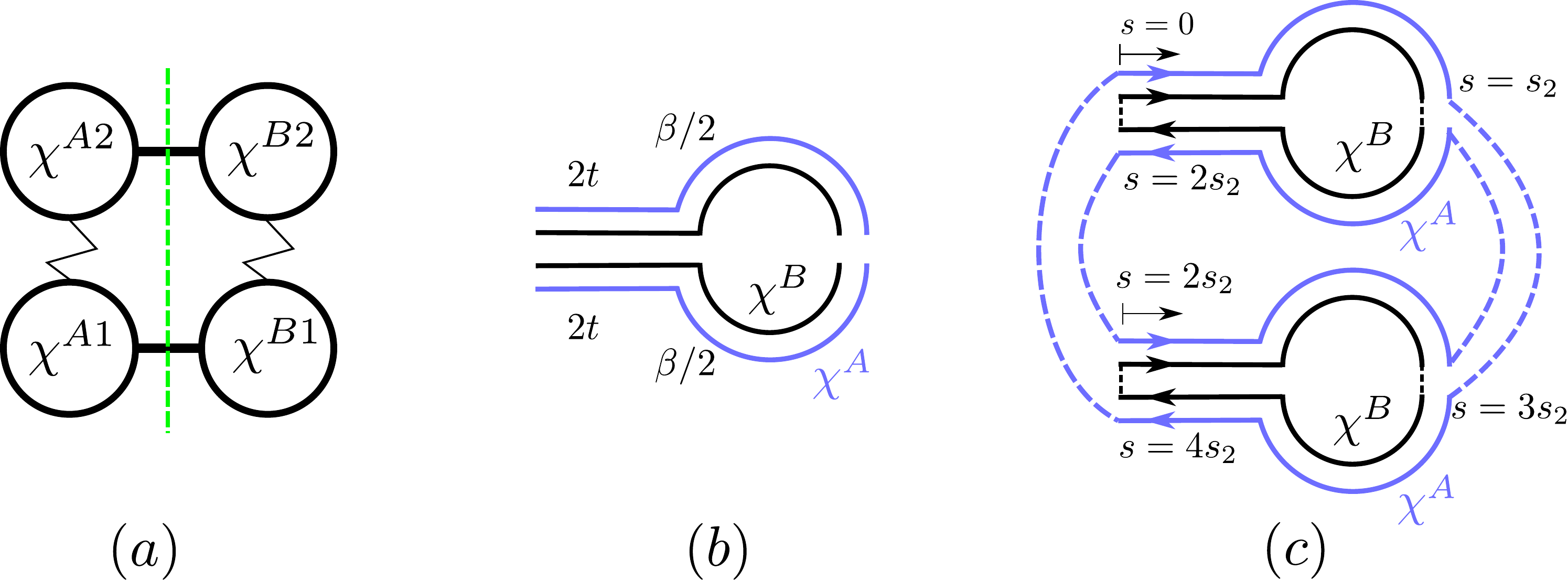}
  \caption{(a) Setup of the TFD quench. The circles represent the different SYK dots and the green dashed line the entanglement cut. 
  (b) Pictorial representation of TFD density matrix. Curved lines indicate imaginary time evolution by $\beta/2$ and horizontal lines real time evolution by $2t$, with the direction from right to left. 
  The lower (upper) curves represent $\ket{TFD_\beta}$ ($\bra{TFD_\beta}$). (c) Contour over which Eq. \eqref{eqn:contour-action} is evaluated. The dashed lines indicate the boundary conditions for the fermions.  
  The contour is parameterized by $s$, which increases clockwise around each replica. 
  Here, $s_2 = 2t+\beta / 2$.}
  \label{fig:tfd-setup}
\end{figure}


As illustrated in Fig. \ref{fig:tfd-setup}(a), we wish to compute the post-quench $n=2$ R\'enyi entropy of the $A1$ and $A2$ fermions, 
\begin{align}
	S^{(2)}_{A1\cup A2} = - \log \mathrm{Tr}_{A1\cup A2} \rho^2_{A1\cup A2}, \label{eqn:tfd-s2-def}
\end{align}
where $\rho_{A1\cup A2}$
is the reduced density matrix for the fermions in region $A1\cup A2$. 
As for the KM state quench, this quantity can be conveniently expressed as a path integral. Indeed, we can pictorially represent the full TFD density matrix, $\rho = \ket{TFD_\beta}\bra{TFD_\beta}$ as in Fig. \ref{fig:tfd-setup}(b). Here, the lower and upper line segments represent $\ket{TFD_\beta}$ and $\bra{TFD_\beta}$, respectively. The curved and horizontal portions of each segment represent the imaginary and forward real time evolutions of the $b=1$ Hilbert space, respectively, read from right to left.  Computing the second R\'enyi involves taking two copies, or replicas, of the density matrix, as in Fig. \ref{fig:tfd-setup}(c). The partial traces over $A$ and $B$ dictate the boundary conditions for $\chi^A_i$ and $\chi_i^B$ fermions, indicated by the dashed lines. 
We can then write the second R\'enyi entropy in terms of a Keldysh path integral over a single copy of the Hilbert space, 
\begin{align}
    \mathrm{Tr}_{A1\cup A2} \rho^2_{A1\cup A2} = \frac{1}{Z(\beta)^2} \int \mathcal{D}\chi_i^A \chi_i^B e^{- I_{\mathcal{C}''} }; \quad
    I_{\mathcal{C}''} = \int_{\mathcal{C}''} ds \left[ \frac{1}{2}\sum_{i,a} \chi_i^a \partial_s \chi_i^a + f(s) H_{q,r} \right] \label{eqn:contour-action} 
\end{align}
where $I_{\mathcal{C}''}$ is
computed over the contour $\mathcal{C}''$ depicted in Fig. \ref{fig:tfd-setup}(c). 
Here, the variable $s$ parameterizes the contour, increasing clockwise around each replica, and takes values in $s \in [0,4s_2)$ where, for brevity, we have defined $s_2 \equiv 2t + \beta/2$. 
Again, $f(s)$ keeps track of whether we are performing imaginary time or forward/backward real time interval at $s$.
Additional details on the setup of this path integral may be found in Appendix \ref{app:discretization}.

On disorder averaging, akin to the KM state calculation, we obtain the effective action
\begin{align}
    \begin{split}
	\frac{I_{\mathcal{C}''}}{N} &=  -\frac{1}{4} \log \det \left(  \underset{a}{\partial_s} - \Sigma_{a}  \right) \frac{1}{4}\int_{\mathcal{C}''} ds ds'  \Bigg[ \sum_{a} \Sigma_{a}(s,s') G_{a}(s,s') \\
	&\qquad\qquad - F(s,s')  \Bigg( \frac{J^2}{q} \sum_{a} G_{a}(s,s')^q + \frac{2V^2}{r} G_A(s,s')^{\frac{r}{2}} G_B(s,s')^{\frac{r}{2}} \Bigg) \Bigg] \end{split}
	\label{eqn:tfd-action}
\end{align}
where $F(s,s') = f(s) f(s')$. 
The saddle-point equations are then given by
\begin{align}
    \begin{split}
	G_a &= (\underset{a}{\partial_s} - \Sigma_{a})^{-1} , \quad 	\Sigma_a(s,s') = F(s,s') \Big[J^2 G_a^{q-1}(s,s')  + V^2 G_a^{r/2-1}(s,s') G_{\bar{a}}^{r/2}(s,s')\Big] .
	\end{split} \label{eqn:tfd-sd-eqns} 
\end{align}
We thus find for the second R\'enyi entropy
    $S^{(2)}_{A1\cup A2}(t) = -I_{\mathcal{C}''} + 2I_\beta$,
where $I_{\mathcal{C}''}$ is evaluated on-shell and $I_\beta = -\log(Z(\beta))$ is the action for a single copy of the coupled SYK model at inverse temperature $\beta$. As in the KM state quench, the Schwinger-Dyson equations can be evaluated numerically. 

\begin{figure}
\subfloat[$V=0.5J$]{%
    \includegraphics[width=0.32\linewidth]{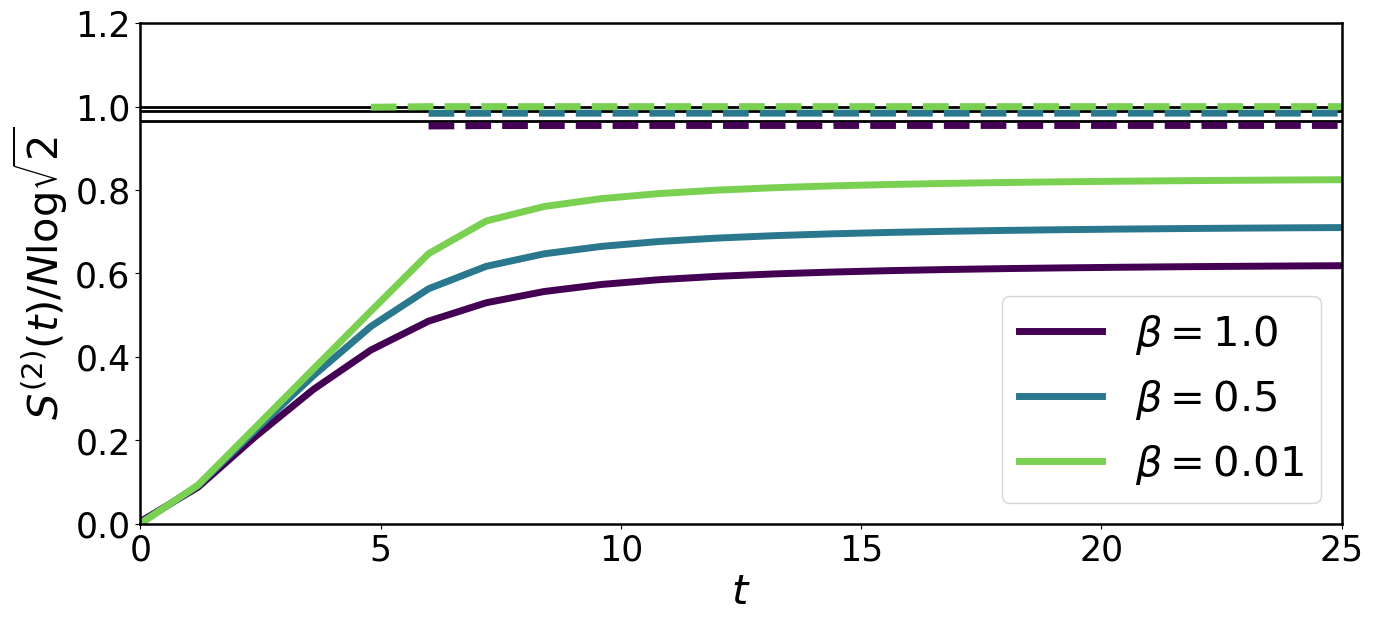}}
\subfloat[$V=J$]{%
    \includegraphics[width=0.32\linewidth]{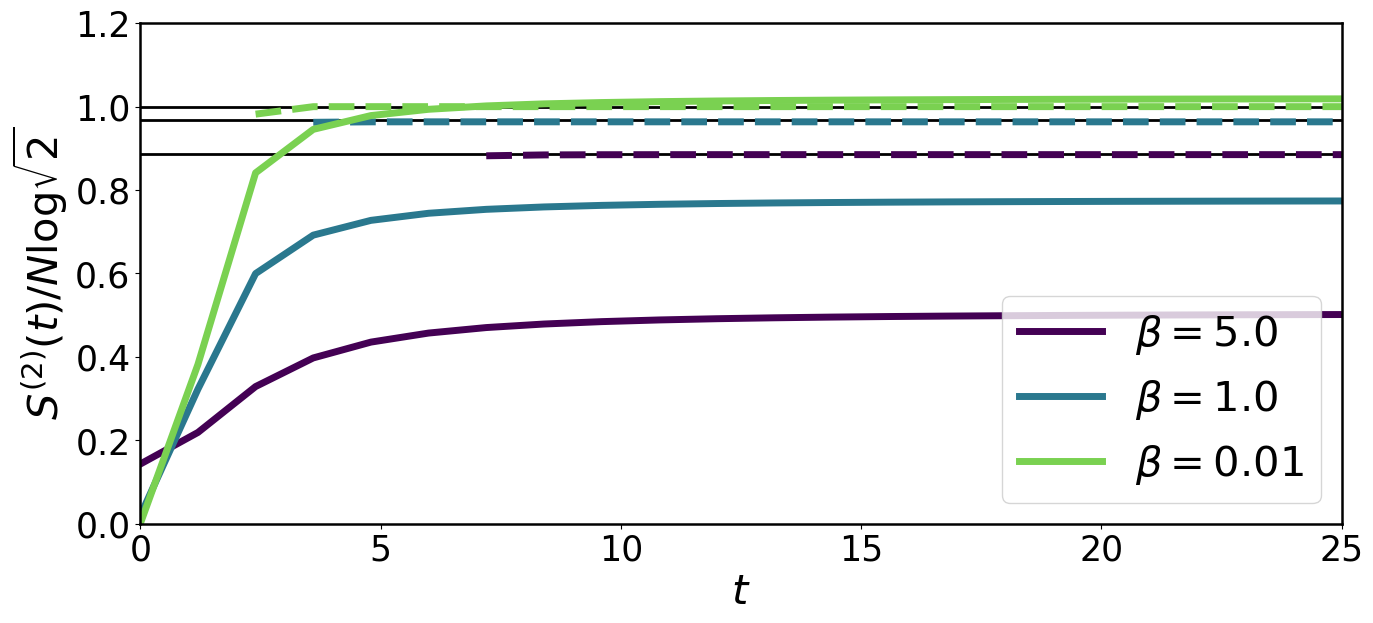}}
\subfloat[$V=1, \, J=0$]{%
    \includegraphics[width=0.32\linewidth]{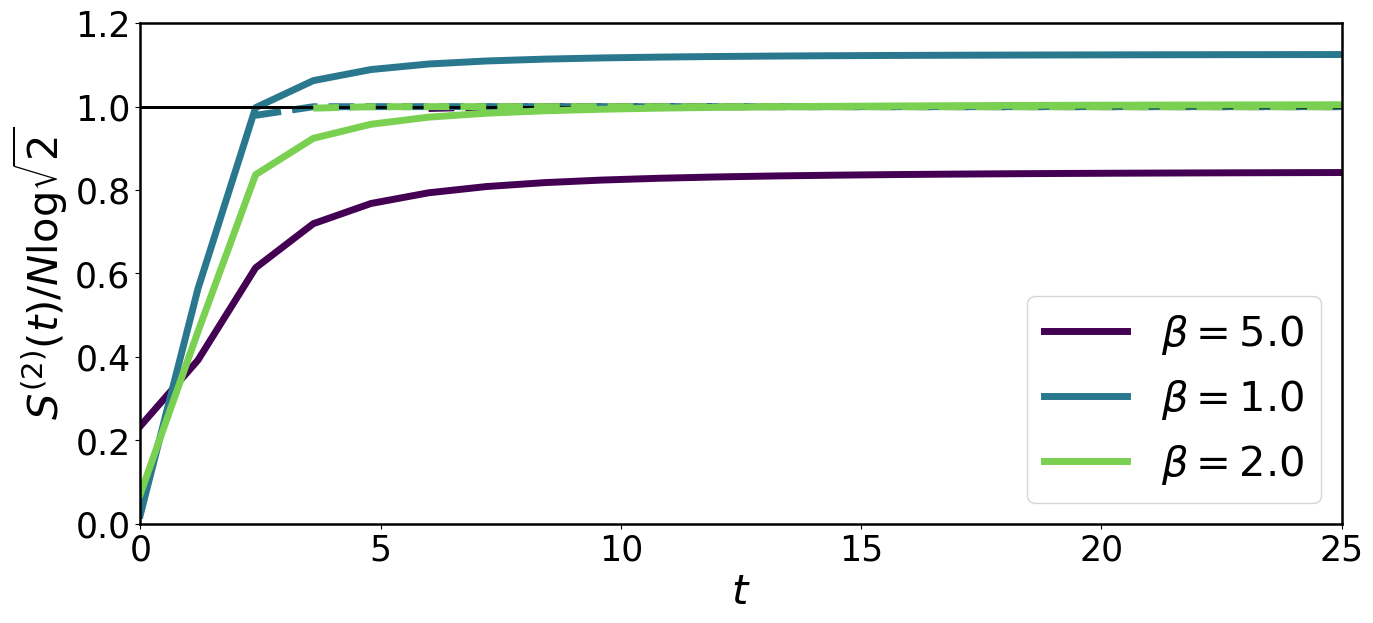}}
  \caption{Second R\'enyi after a quench from different TFD states in the $r=4$ model for different values of $V$ and $J$. The plots follows the same labelling conventions as in Fig. \ref{fig:r4-km-largeN}.  
  } 
  \label{fig:r4-tfd-largeN}
\end{figure}

\begin{figure}
  \subfloat[$V=0.5J$]{%
    \includegraphics[width=0.32\linewidth]{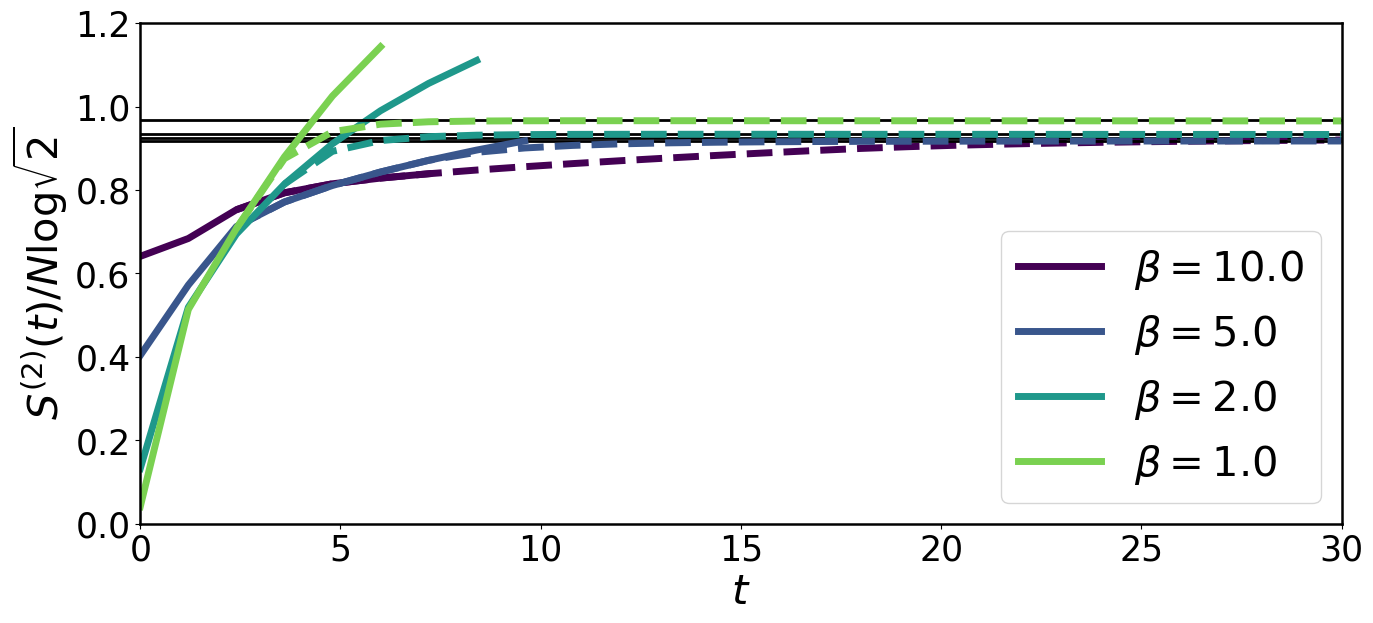}
}%
  \subfloat[$V=0.2J$]{%
    \includegraphics[width=0.32\linewidth]{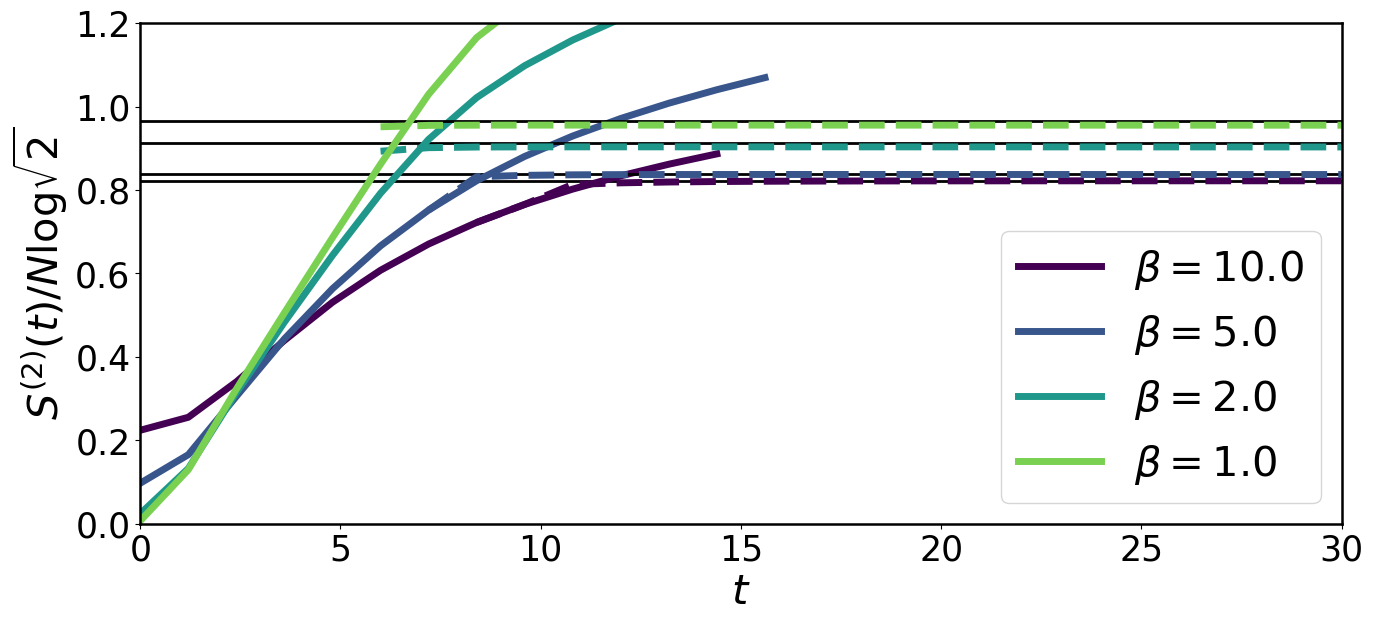}
}
  \subfloat[$V=0.1J$]{%
    \includegraphics[width=0.32\linewidth]{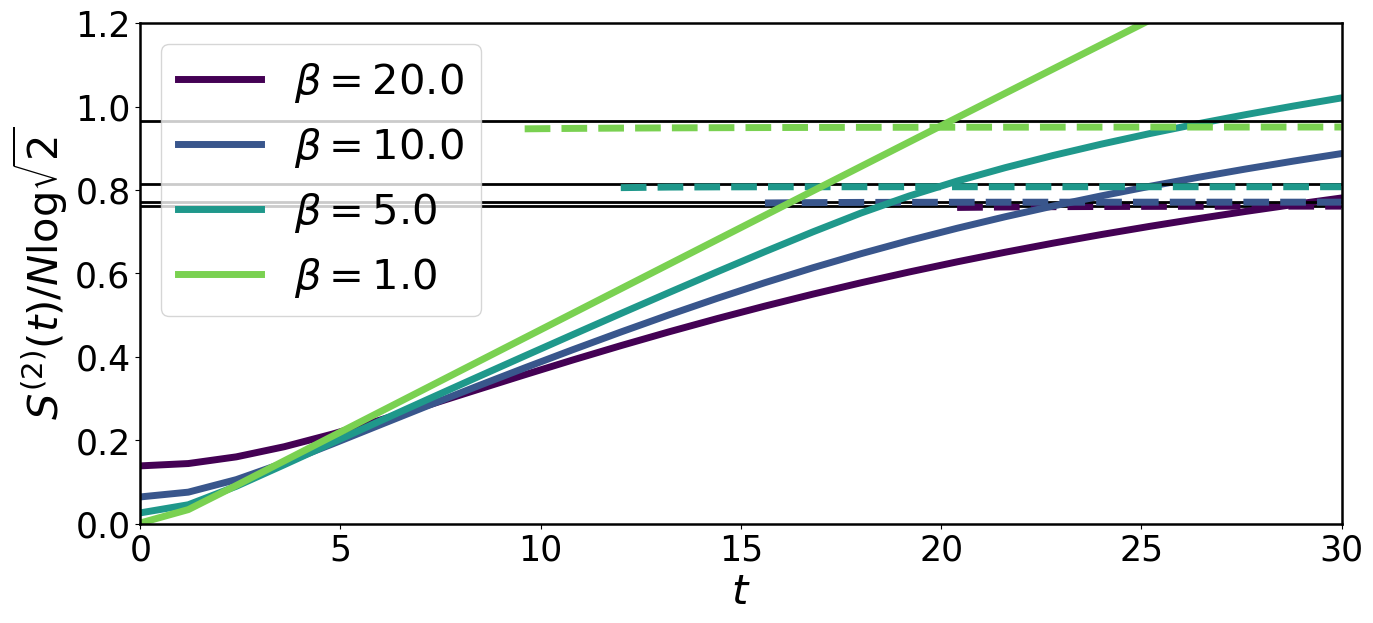}
}
  \caption{ Second R\'enyi after a quench from different TFD states in the $r=2$ model for different values of $V$. 
  }
  \label{fig:r2-tfd-largeN}
\end{figure}

We present the results of our large-$N$ numerics in Fig. \ref{fig:r4-tfd-largeN} for $r=4$, with $V=0.5$ and $V=1$ as well as different values of $\beta$. 
We see that for $V=0.5$, the diagonal solution plateaus to a value below that of the non-diagonal solution, as was noted previously in Ref. \cite{Penington2019}. In fact, our numerics suggest that for sufficiently small values of $V$ (including moderate ones like $V=0.5$), the diagonal saddle-point never crosses the non-diagonal saddle-point, even in the limit $\beta \to 0$, 
and hence the system never thermalizes for $\beta \neq 0$ even though it does for $\beta = 0$.
This non-analytic behavior is quite peculiar and we suspect is an artifact of the large-$N$ limit. Specifically, our results are suggestive of an order-of-limits issue in that taking $\beta \to 0$ and then $N \to \infty$ leads to thermalization whereas taking the limits in the opposite order can lead to subthermal behavior.

As we increase $V$, the plateau value of the replica-diagonal saddle-point increases. 
Indeed, as shown in Fig. \ref{fig:r4-tfd-largeN}(b), for $V=1$, the diagonal saddle-point crosses the non-diagonal saddle-point in $O(1)$ time for $\beta = 0.01$. For this and smaller values of $\beta$, the system is thus in a thermal state at late times. 
But, for larger values of $\beta$, the system remains in a subthermal state. 
In fact, as we see from Fig. \ref{fig:r4-tfd-largeN}(c), even if we set $V=1$ and $J=0$, there exists a critical $\beta$ above which the $r=4$ model does not thermalize. This additional TFD data thus provides additional examples of the state-dependent thermalization of the $r=4$ model. In contrast, from Fig. \ref{fig:r2-tfd-largeN}, we see that the $r=2$ model again appears to always thermalize, as was the case in the KM state quench.

Lastly, we note that one may ask whether the state-dependent thermalization in the $q=4$, $r=2$ model is a consequence of the interdot coupling being more relevant (in the sense of the renormalization group) than the interdot coupling. We have also performed quenches with $q=6$, $r=4$ and again found state-dependent thermalization -- the key point then is whether the interdot coupling is quadratic, not whether it is relevant.

\subsection{Finite-$N$ Quenches}

\begin{figure}
  \subfloat[$r = 4$]{
  \includegraphics[width=0.25\textwidth]{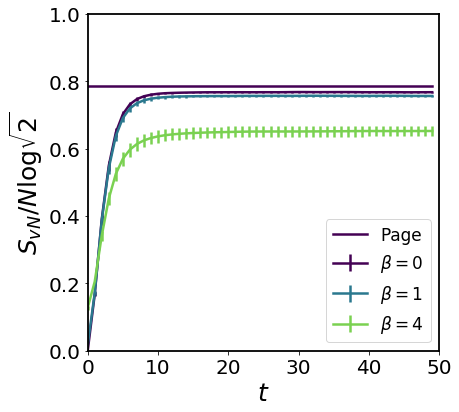}%
  \includegraphics[width=0.25\textwidth]{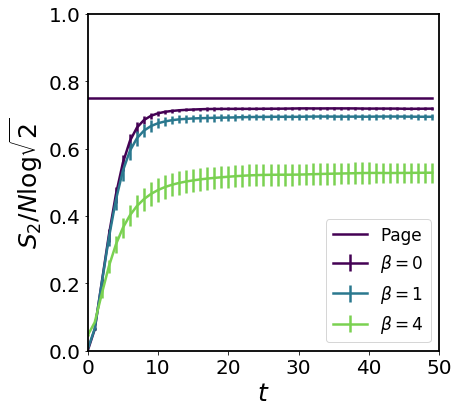}%
}%
  \subfloat[$r = 2$]{
  \includegraphics[width=0.25\textwidth]{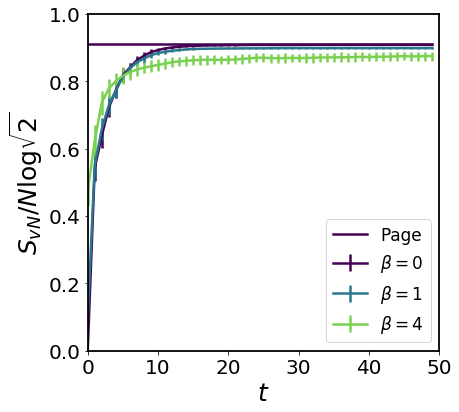}%
  \includegraphics[width=0.25\textwidth]{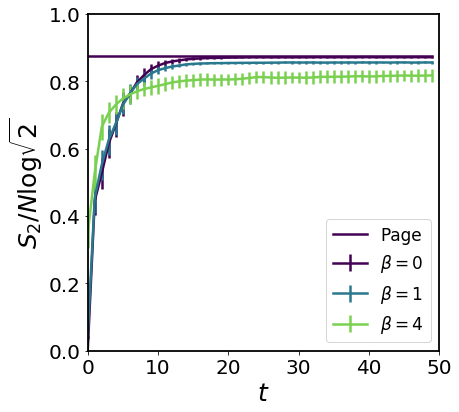}%
}
 \caption{Entanglement and R\'enyi entropies after quenches from TFD states for different values of $\beta$ in the (a) $r=4$ and (b) $r = 2$ coupled models with $N=16$ and $V = 0.5J$, averaged over about 80 disorder realizations. The solid horizontal line in each plot indicates the corresponding Page value of the entropy.
 \label{fig:finiteN-TFD-quench} } 
\end{figure}

We now briefly present data for finite-$N$ quenches. 
Fig. \ref{fig:finiteN-TFD-quench} shows the disorder averaged entanglement and second R\'enyi entropies after a quench from a TFD state in the $r=4$ coupled SYK model with $N = 16$ for various $\beta$. 
The Page values in each plot are determined by Eq.~\eqref{eqn:page-value} in the von Neumann case and Eq.~\eqref{eqn:page-value-renyi} in the R\'enyi case, where $D_A = D_B = 2^{N/2 - 1}$ for $r = 4$ (as the TFD state has definite fermion parity on $A = A_1 \cup A_2$ and on $B = B_1 \cup B_2$), and $D_A = D_B = 2^{N/2}$ for $r = 2$.
Similar to the KM case, in the long time limit the entanglement and R\'enyi entropies from a TFD quench become the same as (or very close to) the Page values. Note that the small discrepancy between the Page values and the long-time TFD values at $\beta = 0$ in the case of $r = 4$ can be attributed to the fact that the eigenstates in the single Hilbert space (which we use to construct the TFD state) have slightly smaller entropies than the corresponding Page values, as shown in Fig.~\ref{fig:estate-entanglement} (b). Finally, it can be numerically shown that there does not exist eigenstates that have the same symmetries as the TFD state, therefore unlike the KM case in Fig.~\ref{fig:finiteN-KM-quench} here we do not have the eigenstate entropies to compare with.

\section{Details of the Large-N Path Integrals \label{app:discretization} }

In this Appendix, we provide additional details on the path integral setup for the KM and TFD state quenches as well as the R\'enyi entropy in the canonical ensemble. 
Here we primarily follow Refs. \cite{Zhang2020a,Liu2020} for the KM state and Ref. \cite{Chen2020} for the TFD state quenches.

\subsection{Kourkoulou-Maldacena State}

\subsubsection{Path Integral Details}

In the main text, we described how to arrive at the contour of Fig. \ref{fig:km-setup}(c) using a pictorial representation of the KM state. 
Here, we provide a more detailed exposition of the path integral setup. Let us first consider the computation of the normalization, $Z_{KM}$. In order to handle the non-trivial boundary conditions of the Majorana fermions given by Eq. \eqref{eqn:km-bcs}, we define a new fermion field, 
\begin{align}
	\tilde{\chi}_i(s) = \begin{cases}
	    -i\chi_{2i-1}^a(\beta/2 - s) \quad & s \in [0 , \beta/2) \\
		\chi_{2i}^a(s-\beta / 2) \quad & s \in [\beta/2,3\beta/2) \\
		-i\chi_{2i-1}^a(5\beta/2 - s) \quad & s \in [3\beta/2 , 2\beta)
	\end{cases}
\end{align}
which satisfies
\begin{align}
    \tchi^a_i(\beta/2^-) &= \tchi^a_i(\beta/2^+), \quad 
    \tchi^a_i(3\beta/2^+) = -\tchi^a_i(3\beta/2^-).
\end{align}
This redefinition allows us to re-express the path integral as one over $N/2$ fermion fields with the usual antiperiodic boundary conditions. 

In order to derive the effective action, we proceed in the usual fashion, first averaging over all disorder realizations, assuming a disorder replica diagonal solution. On inserting a resolution of identity in the form
\begin{align}
    \begin{split}
    1 \propto  \int & \mathcal{D}\Sigma_a \mathcal{D}G_a  \exp \Bigg( - \frac{1}{2} \int ds ds' \sum_a \Sigma_a(s,s') \left[ \frac{N}{4} G_a(s,s') - \sum_i \tchi_i^a(s) \tchi_i^a(s) \right] \Bigg) 
    \end{split} 
\end{align}
into the path integral, we can integrate out the fermions in favor of the bilocal fields $\Sigma_a$ and $G_a$. 
To arrive at Eq. \eqref{eqn:km-action}, we note that the disorder averaged action and boundary conditions are invariant under the simultaneous transformations
    $\chi_{2i-1}^a \mapsto \chi_{2i}^a, \, \chi_{2i}^a \mapsto - \chi_{2i-1}^a,$
which enforces that $G_a(s,s') = G(\beta - s, \beta - s')$ for $s,s' \in [\beta/2,\beta)$ and $G_a(s,s') = G(3\beta - s, 3\beta - s')$ for $s,s' \in [\beta,3\beta / 2)$.
On making use of this symmetry when integrating out the fermions in favor of the bilocal fields $\Sigma_a$ and $G_a$, we may express the effective action in the form of Eq. \eqref{eqn:km-action}. In doing so, it is necessary to introduce the projection function $P(s,s')$ defined in the main text. For the evaluation of $Z_{KM}$, with the parameterization of the contour given in Fig. \ref{fig:km-setup}(b), we have
\begin{align}
    P(s,s') = \begin{cases}
        1  & \!\begin{aligned} &s,s' \in [0,\beta/2) \cup [3\beta/2,2\beta) \, \text{or } s,s' \in [\beta/2,3\beta/2] 
            \!\end{aligned} \\
        0  & \text{otherwise}
    \end{cases} .
\end{align}
Note also that we have $F(s,s') \equiv 1$, as the normalization only involves imaginary time evolution.

The derivation of the effective action for the evaluation of $S^{(2)}_A$, given by Eq. \eqref{eqn:km-action} evaluated on the contour $\mathcal{C}''$ given by Fig. \ref{fig:km-setup}(c), follows in the same vein. Letting $\ket{a}$ and $\ket{a'}$ represent complete bases of states for the $a=A,B$ dots and defining $U = e^{(-it+\beta/2)H}$, we can explicitly write Eq. \eqref{eqn:km-trace} as, 
\begin{align}
    \begin{split}
    Z_{KM}^2 e^{-S^{(2)}_A} =  \sum_{\substack{A,A' , B,B'}} & \braket{\{1\}|U^\dagger |A' B} \braket{A' B'|U|\{1\}} \braket{\{1\}|U^\dagger |A B'} \braket{A B|U|\{1\}} ,
    \end{split}
\end{align}
The graphical representation of this trace is then given by Fig. \ref{fig:km-setup}(c) -- the four overlaps in the above expression correspond to the four half-contours, both read from right to left. The boundary conditions for the $\chi_i^a(\tau)$ fermions indicated by the dashed lines are found by matching up the bras and kets above. The dots in Fig. \ref{fig:km-setup}(c) correspond to boundary conditions of the of form Eq. \eqref{eqn:km-bcs} at $\tau=0,s_1^-,s_1^+, 2s_1$. As in the computation of $Z_{KM}$ these can be accounted for by defining $N/2$ new fermions as
\begin{widetext}
\begin{align}
	\tchi_i(s)^a = \begin{cases}
	    -i\chi_{2i-1}^a(s_1/2 - s +3\alpha s_1) \quad & s \in [2\alpha s_2 , s_1/2 + 2\alpha s_2) \\
		\chi_{2i}^a(s-s_1 / 2  - \alpha s_1) \quad & s \in [s_1/2 + 2\alpha s_2,3s_1/2 + 2\alpha s_2) \\
		-i\chi_{2i-1}^a(5s_1/2 - s +3\alpha s_1) \quad & s \in [3s_1/2 + 2\alpha s_2 , 2s_1 + 2\alpha s_2),
	\end{cases}
\end{align}
\end{widetext}
for $\alpha=0,1$ and where $s \in [0,4s_1)$ reparameterizes the contour, as depicted in Fig. \ref{fig:km-setup}(c). The remaining boundary conditions indicated by the dashed lines are, explicitly,
\begin{align}
\begin{split}
    \tchi^A(2s_1^-) &= \tchi^A(0^+), \, \tchi^A(s_1^-) = \tchi^A(s_1^+), \\
    \tchi^A(4s_1^-) &= \tchi^A(2s_1^+), \, \tchi^A(3s_1^-) = \tchi^A(3s_1^+);
    \end{split}
    \begin{split}
    \tchi^B_i(0^+) &= \tchi^B_i(4s_1^-), \, \tchi^B_i(3s_1^+) = \tchi^B_i(s_1^-), \\
    \tchi^B_i(2s_1^+) &= \tchi^B_i(2s_1^-), \, \tchi^B_i(3s_1^-) = \tchi^B_i(s_1^+).
    \end{split}
\end{align}
On performing the disorder average, we again arrive at the effective action Eq. \eqref{eqn:km-action}, now evaluated on the contour $\mathcal{C}''$. Here, from inspection of Fig. \ref{fig:km-setup}, we have that
\begin{align}
    P(s,s') = \begin{cases}
        1 & \!\begin{aligned}  & s,s' \in [0,s_1/2]\cup[3s_1/2,5s_1/2]\cup[7s_1/2,4s_1] \, \text{or } s,s' \in [s_1/2,3s_1/2]\cup [5s_1/2,7s_1/2],
        \!\end{aligned} \\
        0 & \text{otherwise}
    \end{cases}
\end{align}
where $s_1 \equiv 2t+\beta$. Since we now have real time evolution, $F(s,s') = f(s) f(s')$ is non-trivial. Explicitly,
\begin{align}
    f(s) = \begin{cases}
    1 & s \in [t,t+\beta) \cup [\beta+3t,2\beta+3t) \\
    -i & s \in [0,t)\cup [2\beta+3t,2\beta+4t) \\
    i & s \in [\beta+t,\beta+3t)
    \end{cases}.
\end{align}
For $s \in [2s_1,4s_1)$, we set $f(s) = f(s-2s_1)$.

Now, when we discretize the path integral, the distinct boundary conditions for the Majorana fermions will be encoded in the matrix used to define the time derivative operator. 
Following Ref. \cite{Chen2020}, we employ the fact that $\underset{a}{\partial_s} = (G_a^0)^{-1}$,
where $G^0_a(s,s')$ is the free propagator with the appropriate boundary conditions for the $a=A,B$ fermions. 
In the evaluation of $Z_{KM}$, the $\tchi^a$ fermions obey the same standard fermion boundary conditions, and so 
\begin{align}
    G_a^0(s,s') = \frac{1}{2} \sgn(s-s'), \quad a= A,B.
\end{align}
In the evaluation of $S^{(2)}_A$, we instead have distinct boundary conditions for the $a=A,B$ fermions; explicitly,
\begin{align}
    G_{a}^0(s,s') = \begin{cases}
    \frac{1}{2} \sgn(s-s') \quad &s,s' \text{ on the same } a \text{ contour} \\
    0 \quad &\text{otherwise} .
    \end{cases} \label{eqn:free-green-fn-renyi}
\end{align}
For instance, we can read off from Fig. \ref{fig:km-setup}(c) that $G_{A}^0(t/2,7t+4\beta) = 0$, while $G_{B}^0(t/2,7t+4\beta) = -1/2$.

\subsubsection{Numerical Evaluation}

In order to numerically solve the Schwinger-Dyson equations, Eq. \eqref{eqn:km-sd-eqns}, and evaluate the action, Eq. \eqref{eqn:km-action}, we discretize the contour into into $L$ points, turning the Green functions and self-energies into $L \times L$ matrices. In particular, we divide each real time interval
[e.g. $s \in [0,t)$] into $T$ points and each imaginary time interval [e.g. $s \in (t,t+\beta)$] into $B$ points, so that $L = 8T + 4B$. 
After discretization, we have that $ds \to \Delta s_i$, where $\Delta s_i = t/T$ or $\Delta s_i = \beta/B$,  depending on whether $s_i$ is in a real or imaginary time interval. 
Additionally, we have
\begin{align}
    \begin{split}
	\delta(s-s')\underset{a}{\partial_s} &\to \frac{1}{\Delta s_m \Delta s_n}(G_{a}^0)^{-1}_{mn}, \qquad 
	\Sigma_a(s,s') \to \frac{1}{\Delta s_m \Delta s_n} (\Sigma_a)_{mn}, 
	\end{split}
\end{align}
where additional factors of $\Delta s_m$ are included in the discretization of $\partial_s$ to retain the correct units, while the rescaling of $\Sigma_a$ is done for convenience.
The discretized Schwinger-Dyson equations then take the form
\begin{align}
	(G_a)_{mn} &= (\underset{a}{\partial_s} - \Sigma_{a})^{-1}_{mn} , \qquad  (\Sigma_a)_{mn} = P_{mn} F_{mn} \left[J^2 G_a^{q-1} + V^2 G_a^{r/2-1} G_{\bar{a}}^{r/2}\right]_{mn} \Delta s_m \Delta s_n.
\end{align}
We solve the Schwinger-Dyson equations using a standard self-consistent iterative procedure. Starting with some \textit{ansatz\"e} for the Green functions on iteration $l=1$, $G_a^{(l=0)}$, we compute the self-energies $[\Sigma_a]^{(l)}$ using the Schwinger-Dyson equations and perform a weighted update of the Green functions:
\begin{align}
    G_a^{(l+1)} = (1-x)G_a^{(l)} + x \left( (G_{a}^0)^{-1} - \Sigma_a^{(l)} \right)^{-1} ,
\end{align}
where we typically take $x=0.5$. The Green functions of the $l^{th}$ iteration are then used as input for the subsequent iteration. We say the procedure has converged once
the change in Green functions between iterations
    $\sum_{a,m,n} | (G_a^{(l)})_{mn} - (G_a^{(l)})_{mn}|/(2L^2)$
drops below a threshold of $\epsilon = 10^{-9}$ and the Schwinger-Dyson equations are satisfied to the same tolerance. We then use these values to compute the action, Eq. \eqref{eqn:tfd-action}. We repeat this computation for several values of $L$ and then perform a linear extrapolation in $1/L$ to extract the $L \to \infty$ value. As our initial \textit{ansatz\"e} for the Green functions, we set either $G_a^{l=0} = G_a^0$ or $G_a^{l=0} = G_B^0$; the former tends to converge to the replica-diagonal saddle-point and the latter to the non-diagonal saddle-point.

\subsection{Thermofield Double State}

Next, we briefly outline the setup of the contour of Fig. \ref{fig:tfd-setup}(c) starting from
Eq. \eqref{eqn:tfd-s2-def}.
Let us first define, for notational convenience, $U_1 \equiv e^{-(i2t+\beta/2)H_1}$. 
Now, since $\ket{I}_{12}$ is an infinite temperature TFD state (i.e. a product state of Bell pairs between the $b=1$ and $b=2$ Hilbert spaces, as indicated by Eq. \eqref{eqn:tfd-inf-temp-cnd}), we have that  $\Tr_{B2} (\ket{I}_{12} \prescript{}{12}{\bra{I}}) = \ket{I}_{A1\cup A2} \prescript{}{A1\cup A2}{\bra{I}} \otimes \mathbf{1}_{B1}$, where $\ket{I}_{A1\cup A2} \prescript{}{A1\cup A2}{\bra{I}}$ is the infinite temperature TFD state for $A1\cup A2$ and $\mathbf{1}_{B1}$ the identity on $B1$. Making use of this fact and inserting complete bases of states in the trace, we find 
\begin{align}
    \begin{split}
	Z(\beta)^2 e^{-S^{(2)}_A} = \sum_{\substack{A_1, A_1' , \tilde{A}_1 \tilde{A}_1'  \\ B_1, B_1' , 
	\tilde{B}_1 , \tilde{B}'_1 }}  \bra{\tilde{A}_1  \tilde{B}'_1}  U_1^\dagger \ket{A_1 B_1'} \bra{A_1' B_1'} U_1 \ket{\tilde{A}_1' \tilde{B}_1'} \bra{\tilde{A}_1 ' \tilde{B}_1}  U_1^\dagger \ket{A_1' B_1} \bra{A_1 B_1} U_1 \ket{\tilde{A}_1 \tilde{B}_1} ,
	\end{split}
	\label{eqn:overlaps}
\end{align}
Here, $\ket{a_1}, \, \ket{a'_1}$, and $\ket{\tilde{a}_1}$ represent complete bases of states for the $a=A,B$ dots in the $b=1$ Hilbert space.
Note that the $b=2$ Hilbert space has been traced out. The second R\'enyi entropy thus has a Keldysh path integral representation given by Eq. \eqref{eqn:contour-action}.  Indeed, read from left to right, the four overlaps in Eq. \eqref{eqn:overlaps} correspond to the four contour segments in Fig. \ref{fig:tfd-setup}(c), read from top to bottom. The boundary conditions for the fermions, indicated by the dashed lines connecting the segments, can be deduced by matching up the bras and kets in Eq. \eqref{eqn:overlaps}, as we described for the KM quench. The free Green functions for the $a=A,B$ fermions again takes the form of Eq. \eqref{eqn:free-green-fn-renyi}.

\subsection{R\'enyi Entropy in Thermal Ensemble}

Finally, we outline the computation of the R\'enyi entropy between the $A$ and $B$ dots in the Gibbs state $\rho_\beta = e^{-\beta H} / Z(\beta)$, against which we compared the large-$N$  non-diagonal saddle-points, following Ref. \cite{Zhang2020a}. Explicitly,
\begin{align}
    S_A^{(2)} = \frac{1}{Z(\beta)^{2}} \sum_{A,A',B,B'} \braket{AB|e^{-\beta H}|A'B} \braket{A'B'|e^{-\beta H}|AB'}.
\end{align}
Expressed as a path integral, the contour takes the standard ``pants" geometry. The replica boundary conditions implied by the above expressions yield for the free Green functions,
\begin{align}
    G^0_A(\tau,\tau') &= \frac{1}{2}\sgn(\tau-\tau') \,\, \tau,\tau' \in [0,2\beta), \quad     \notag G^0_B(\tau,\tau') = \frac{1}{2}\sgn(\tau-\tau') \,\, \tau,\tau' \in [0,\beta)\text{ or } \tau,\tau' \in [\beta,2\beta), 
\end{align}
where $\tau \in [0,2\beta)$. The R\'enyi entropy in the Gibbs state can then be evaluated by using these free Green functions in Eq. \eqref{eqn:tfd-action} and solving the saddle-point equations, as in the preceding two subsections.

\bibliography{references}

\end{document}